\documentclass[12pt]{article}

\title{Kinetic theory of two dimensional point vortices  \\
 from a BBGKY-like hierarchy}

\usepackage{amsthm,amsmath,amssymb,a4wide}
\usepackage{color}

cm \hsize=18 true cm \topmargin=-2cm \oddsidemargin=-0.5cm
\def\mb#1{\setbox0=\hbox{$#1$}\kern-.025em\copy0\kern-\wd0
\kern-0.05em\copy0\kern-\wd0\kern-.025em\raise.0233em\box0}

\begin{document}

\author{Pierre-Henri Chavanis}
\maketitle
\begin{center}
Laboratoire de Physique Th\'eorique (CNRS UMR 5152), \\
Universit\'e
Paul Sabatier,\\ 118, route de Narbonne, 31062 Toulouse Cedex 4, France\\
E-mail: {\it chavanis{@}irsamc.ups-tlse.fr\\
 }
\vspace{0.5cm}
\end{center}

\begin{abstract}

Starting from the Liouville equation, we derive the exact hierarchy of
equations satisfied by the reduced distribution functions of the
single species point vortex gas in two dimensions. Considering an
expansion of the solutions in powers of $1/N$ (where
$N$ is the number of vortices) in a proper thermodynamic limit
$N\rightarrow +\infty$, and neglecting some collective effects, we
derive a kinetic equation satisfied by the smooth vorticity field
which is valid at order $O(1/N)$.  This equation was obtained
previously [P.H. Chavanis, Phys. Rev. E, 64, 026309 (2001)] from a
more abstract projection operator formalism.  If we consider
axisymmetric flows and make a markovian approximation, we obtain a
simpler kinetic equation which can be studied in great detail. We
discuss the properties of these kinetic equations in regard to the
$H$-theorem and the convergence (or not) towards the statistical
equilibrium state. We also study the growth of correlations by
explicitly calculating the time evolution of the two-body correlation
function in the linear regime.  In a second part of the paper, we
consider the relaxation of a test vortex in a bath of field vortices
and obtain the Fokker-Planck equation by directly calculating the
second (diffusion) and first (drift) moments of the increment of
position of the test vortex. A specificity of our approach is to
obtain general equations, with a clear physical meaning, that are
valid for flows that are not necessarily axisymmetric and that take
into account non-Markovian effects. A limitations of our approach,
however, is that it ignores collective effects.

\vskip1cm

\end{abstract}
\eject

\section{Introduction}
\label{sec_introduction}

Several authors have wondered whether fluid turbulence could be
described in terms of statistical mechanics
\cite{frisch}. Three dimensional 
turbulence has been attacked by different methods
\cite{kivotides,she,beck,monchaux} inspired by statistical mechanics and kinetic theories. Some progress has also been made in the simpler
case of two dimensional turbulence (see reviews in
\cite{sommeria,tabeling,houches}). Two dimensional turbulence is not
just academic but is relevant to describe geophysical and
astrophysical flows.  Two dimensional flows are characterized by the
spontaneous formation of large-scale vortices that dominate the
dynamics
\cite{mw,bracco}. The most famous example is Jupiter's great
red spot, a huge vortex persisting for more than three centuries in a
turbulent shear layer between two zonal jets in the southern
hemisphere of the planet \cite{marcus}.  Other examples of this
self-organization are the cyclones and anticyclones in the earth
atmosphere, the jets in the oceans like the gulf stream or the intense
jets on Jupiter \cite{pedlosky}. As a first step to tackle the
problem, it can be of interest to study the dynamics of a system of
$N$ point vortices on a plane \cite{newton}.  Each vortex produces a
velocity field that moves the other vortices in a self-consistent
manner. The velocity created by a vortex decreases like $1/r$ which is
similar to the Coulombian or Newtonian interaction in two
dimensions. Therefore, the interaction between point vortices is
long-range, like the interaction between stars in a galaxy or between
electric charges in a plasma. Note, however, that point vortices
produce a velocity while material particles produce a force
(acceleration). Apart from this (important) difference, the point
vortex gas has a Hamiltonian structure \cite{kirchhoff} and we can try
to apply the methods of statistical mechanics and kinetic theory to
that system. Therefore, the $N$-vortex problem \cite{newton} is of
fundamental interest in statistical mechanics and kinetic theory. It
provides a physical example of systems with long-range interactions,
whose dynamics and thermodynamics are actively studied at present
\cite{dauxois}.

The statistical mechanics of 2D point vortices was first considered by
Onsager \cite{onsager} in a seminal paper. He showed that statistical
equilibrium states with sufficiently large energies have negative
temperatures. For such states, like-sign vortices have the tendency to
group themselves and form clusters. If the circulations of all the
point vortices have the same sign, the equilibrium state is a
large-scale vortex (supervortex) similar to vortices observed in
geophysical and astrophysical flows.  When the point vortices have
positive and negative circulations, the equilibrium state is
generically a dipole made of a cluster of positive vortices and a
cluster of negative vortices. The pioneering work of Onsager was
pursued by Joyce \& Montgomery \cite{jm} and Lundgren \& Pointin
\cite{lp}, using a mean field approximation. Using a combinatorial
analysis, Joyce \& Montgomery introduced an entropy for the point
vortex gas which is similar to the Boltzmann entropy for material
particles. The statistical equilibrium state (most probable) is
obtained by maximizing this Boltzmann entropy while conserving all the
constraints imposed by the dynamics (total number $N$ of point
vortices and energy $E$, as well as angular momentum $L$ and impulse
${\bf P}$ for domains with a special symmetry). For point vortices with
equal circulation $\gamma$, the smooth vorticity field is given by the
Boltzmann distribution $\omega({\bf r})=A e^{-\beta\gamma\psi({\bf
r})}$, where the potential is played by the stream function $\psi({\bf
r})$. Using $\omega=-\Delta\psi$, the stream-function is then
determined by the Boltzmann-Poisson equation.  Lundgren
\& Pointin started from the exact equilibrium hierarchy of equations
satisfied by the reduced distribution functions $P_j({\bf
r}_1,...,{\bf r}_j)$ of the point vortex gas and, by neglecting all
the correlations between point vortices, derived a differential
equation determining the equilibrium distribution of the one-body
distribution function $P_1({\bf r}_1)$.  Using the fact that
$\omega({\bf r})=N\gamma P_1({\bf r})$, the mean field equation
derived by Lundgren \& Pointin coincides with the Boltzmann-Poisson
equation derived by Joyce \& Montgomery. In a mathematical work,
Caglioti {\it et al.} \cite{caglioti} showed rigorously that the mean
field approximation is exact in a proper thermodynamic limit
$N\rightarrow +\infty$ such that $\gamma\sim 1/N$, $E\sim 1$,
$\beta\sim N$ and $V\sim 1$ (where $V$ is the area of the domain). In
that limit the $N$-body distribution at statistical equilibrium is a
product $P_{N}({\bf r}_{1},...{\bf r}_{N})=P_{1}({\bf r}_{1})
... P_{1}({\bf r}_{N})$ of $N$ one-body distributions that are
solution of the Boltzmann-Poisson equation. This statistical
equilibrium state is expected to be achieved for $t\rightarrow
+\infty$. We stress, however, that the statistical theory is based on
the assumption that ``at statistical equilibrium, all accessible
microstates are equiprobable''. This is essentially a postulate, so
there is no guarantee that the point vortex gas will reach a
statistical equilibrium state of the form described above (based on
the microcanonical distribution). In order to determine the timescale
of the relaxation of the smooth vorticity field $\omega({\bf r},t)$,
and in order to establish whether (or not) the system will truly relax
towards Boltzmann statistical equilibrium, we must develop a
kinetic theory of point vortices.

A kinetic theory was developed by Dubin \& O'Neil \cite{dn} in the
case of a non neutral plasma confined by a strong magnetic field, a
system isomorphic to the point vortex gas. They started from the
Klimontovich equation and used a quasilinear approximation to
determine the current of the smooth density due to discrete
interactions between point vortices. They considered an axisymmetric
evolution of the system and, in the course of their derivation, made a
Markov approximation assuming that the two-body correlation function
relaxes on a timescale that is much shorter than the timescale on
which the smooth density field changes (this is the counterpart of the
Bogoliubov hypothesis in plasma physics). They obtained a closed
expression of the current, see Eq. (11) of \cite{dn}, taking into
account ``collective effects'' between the particles. These collective
effects are similar to those giving rise to the Debye shielding in
plasma physics in the Lenard-Balescu approach
\cite{lenard,balescu}. In plasma physics, they take into account the
fact that a charge is surrounded by a polarization cloud of opposite
charges.  In the case of point vortices, their physical interpretation
and their consequence is more difficult to establish.

A kinetic theory of point vortices was carried out independently by
Chavanis \cite{kin}, using an analogy with the kinetic theory
developed for stellar systems. He started from the Liouville equation
and used the projection operator formalism of Willis \& Picard
\cite{wp} to derive a kinetic equation for the smooth vorticity
distribution $\omega({\bf r},t)$. By this method, he obtained a
kinetic equation, see Eq. (128) of \cite{kin}, that is valid for flows
with arbitrary symmetry (non necessarily axisymmetric) and taking into
account memory effects. This is the counterpart of the generalized
Landau equation in stellar dynamics derived by Kandrup \cite{kandrup1}
using the same formalism, see Eq. (42) of \cite{kandrup1}. If we
restrict ourselves to axisymmetric flows and make a Markovian
approximation, this leads to a simplified kinetic equation, see Eq.
(133) of \cite{kin}, which coincides with the equation obtained by
Dubin \& O'Neil \cite{dn} when collective terms are ignored. In a
sense, the simplified kinetic equation (133) obtained by Chavanis
\cite{kin} is the counterpart of the Landau \cite{landau} 
equation in plasma physics
while the more general equation (11) obtained by Dubin
\& O'Neil \cite{dn} is the counterpart of the Lenard-Balescu equation \cite{lenard,balescu}.  
Finally, the general non-Markovian equation (128) of \cite{kin} is
related to the Master equations discussed by Prigogine
\cite{prigogine} in plasma physics. Therefore, there are many
interesting analogies between plasma physics, stellar dynamics and
vortex dynamics. In these analogies, the position ${\bf r}$ of the
point vortices plays the role of the velocity ${\bf v}$ of the
particles in a plasma or in a stellar system, and the angular momentum
$L=\int
\omega r^{2} d{\bf r}$ plays the role of the kinetic energy
$K=\frac{1}{2}\int f v^{2}d{\bf v}$. Chavanis \cite{kin} also
considered the evolution of a test vortex in a bath of field vortices
with fixed distribution (e.g. a thermal bath at statistical
equilibrium) and used the projection operator formalism to derive a
Fokker-Planck equation for the evolution of the one-body distribution
of the test vortex $P({r},t)$ in the bath. This equation involves a
term of diffusion and a term of drift that are both position
dependent. For a thermal bath, i.e. when the field vortices are at
statistical equilibrium, the diffusion coefficient and the drift
coefficient (mobility) are related to each other by a sort of Einstein
relation involving a negative temperature (in cases of physical
interest). The resulting Fokker-Planck equation, see Eq. (115) of
\cite{kin}, is the counterpart of the Kramers-Chandrasekhar equation, see
Eq. (10) of \cite{kc}, in stellar dynamics.

The projection operator formalism which is at the basis of the
above-mentioned kinetic theory is very abstract and it is not clear to
understand which approximations have been made in the course of the
derivation. In this paper, we present an alternative derivation of the
kinetic equations obtained in \cite{kin} which is more transparent. We
start from the exact out-of-equilibrium BBGKY-like hierarchy of equations
satisfied by the reduced distribution functions $P_j({\bf
r}_1,...,{\bf r}_j,t)$ of the point vortex gas and consider an
expansion of the solutions in powers of $1/N$ in a proper thermodynamic
limit $N\rightarrow +\infty$. The kinetic equations obtained in
\cite{kin} are recovered at the order $O(1/N)$. The idea of 
using a BBGKY-like  hierarchy to derive the kinetic
equations of \cite{kin} was given in \cite{chavlemou}. This
derivation has the advantage of being much simpler than the previous
one and shows clearly the domain of validity of the kinetic
equations. It also shows how collective terms can be included in the
calculations. However, we shall not try to evaluate these terms in the
present paper because they require heavy calculations in the complex
plane similar to those performed in plasma physics to derive the
Lenard-Balescu equation from the BBGKY hierarchy. We hope to come to
this problem in a future work.

The paper is organized as follow. In Sec. \ref{sec_ses}, we consider
the statistical equilibrium state. We recall the equilibrium hierarchy
derived by Lundgren \& Pointin \cite{lp} and the proper thermodynamic
limit for the point vortex gas.  For $N\rightarrow +\infty$, the
two-body correlation function vanishes so that the mean field
approximation is exact in that limit. We derive the differential
equation satisfied by the smooth vorticity field. For the usual
potential of interaction between point vortices, it reduces to the
Boltzmann-Poisson equation. We go beyond the mean field approximation
and determine the exact integrodifferential equation satisfied by the
two-body correlation function at order $O(1/N)$. In
Sec. \ref{sec_bbgky}, we consider the out-of-equilibrium problem and
derive the BBGKY-like hierarchy satisfied by the reduced distribution
functions of the single species point vortex gas. We close the
hierarchy by considering an expansion of the solutions in powers of
$1/N$ in the proper thermodynamic limit $N\rightarrow +\infty$
(Sec. \ref{sec_h}). To leading order in $N\rightarrow +\infty$, we
find that the smooth vorticity field satisfies the 2D Euler
equation. This is the counterpart of the Vlasov equation in plasma
physics and stellar dynamics. At order $O(1/N)$, we obtain an exact
system of coupled equations taking into account ``distant collisions''
between point vortices. If we neglect collective effects, we can
obtain an explicit kinetic equation that is valid for flows that are
not necessarily axisymmetric and that takes into account non-markovian
effects. It is valid on a timescale $\sim N t_{D}$. We simplify this
equation by considering axisymmetric flows and arguing that markovian
effects can be neglected for $N\rightarrow +\infty$
(Sec. \ref{sec_ma}). We discuss the properties of these kinetic
equations in regard to the $H$-theorem and the convergence (or not)
towards the statistical equilibrium state (Sec.
\ref{sec_disc}). We also study the growth of correlations by explicitly
calculating the time evolution of the two-body correlation function in
the linear regime (Sec. \ref{sec_g}). In Sec. \ref{sec_q}, we
show that the above-mentioned kinetic equations can also be obtained
from a quasilinear theory starting from the Klimontovich
equation. This is the approach considered by Dubin \& O'Neil \cite{dn}
for axisymmetric flows. We show how it can be generalized to arbitrary
flows when collective effects are neglected. We also
stress the connection with the quasilinear theory of the 2D
Euler-Poisson system developed by Chavanis \cite{prl} to describe the
process of violent relaxation in the collisionless regime
\cite{lb,miller,rs,csr}.  Finally, in Sec. \ref{sec_r}, we consider 
the relaxation of a test vortex in a bath of field vortices at
equilibrium and obtain the Fokker-Planck equation by directly
calculating the second (diffusion) and first (drift) moments of the
increment of position of the test vortex. We obtain general
expressions, with a clear physical meaning, that are valid for flows
that are not necessarily axisymmetric and that take into account
non-Markovian effects. We simplify them in the case of axisymmetric
flows and recover by a direct calculation the Fokker-Planck equation
that was obtained in \cite{kin} from the more formal projection
operator formalism. We also indicate how the results can be
generalized when collective effects are taken into account in the
problem.

\section{The statistical equilibrium state}
\label{sec_ses}

To set the notations and show the connection with the kinetic theory
developed in the next section based on a BBGKY-like hierarchy, we
first derive the differential equation satisfied by the smooth
vorticity profile at statistical equilibrium from an equilibrium  hierarchy
 \cite{lp}.

\subsection{The equilibrium hierarchy}
\label{sec_b}

The exact vorticity field of a gas of point vortices with equal
circulation $\gamma$ is given by
\begin{eqnarray}
\label{b1}
\omega_{d}({\bf r},t)=\sum_{i=1}^{N}\gamma \delta({\bf r}-{\bf
r}_{i}(t)),
\end{eqnarray}
where ${\bf r}_{i}(t)$ is the position of the $i$-th point vortex at
time $t$. The dynamical evolution of the point vortices is governed
by the Hamilton equations
\begin{eqnarray}
\label{b2} \gamma{d{x}_{i}\over dt}={\partial H\over\partial {y}_{i}},
\qquad \gamma{d{y}_{i}\over dt}=-{\partial H\over\partial {x}_{i}}, \nonumber\\
H=\gamma^{2}\sum_{i<j}u(|{\bf r}_{i}-{\bf r}_{j}|),\qquad
\end{eqnarray}
where the positions $(x,y)$ of the point vortices are canonically
conjugate \cite{kirchhoff}. For simplicity, we present the results
in an unbounded domain where the potential of interaction depends
only on the absolute distance between point vortices, but we stress
that most of our results can be extended to bounded domains by using
the generalized Green function of Lin \cite{lin} satisfying the reciprocity
property $u({\bf r}_{i},{\bf r}_{j})=u({\bf r}_{j},{\bf r}_{i})$.
The $N$-vortex distribution function satisfies the Liouville
equation
\begin{eqnarray}
\label{b3}
\frac{\partial P_N}{\partial t}+\sum_{i=1}^{N}{\bf
V}_{i}\frac{\partial P_N}{\partial {\bf r}_i}=0,
\end{eqnarray}
expressing the conservation of the probability density $P_N({\bf
r}_{1},...,{\bf r}_{N},t)$ that the first point vortex is in ${\bf
r}_{1}$, the second in ${\bf r}_{2}$ etc., at time $t$ and where ${\bf
V}_{i}$ is the total velocity of point vortex $i$. It is created by
the other vortices, so that
\begin{eqnarray}
\label{b4}
{\bf V}_{i}=\sum_{j\neq i}{\bf V}(j\rightarrow i),
\end{eqnarray}
where
\begin{eqnarray}
\label{b5}
{\bf V}(j\rightarrow i)=-\gamma {\bf z}\times \frac{\partial
u_{ij}}{\partial {\bf r}_i},
\end{eqnarray}
is the velocity created by point vortex $j$ on point vortex $i$. We
shall essentially consider the standard potential of interaction
$u_{ij}=-(1/2\pi)\ln |{\bf r}_i-{\bf r}_j|$ but we leave the function
$u(|{\bf r}_{i}-{\bf r}_{j}|)$ as general as possible in order to
describe different models like, e.g., the quasi-geostrophic model. Any
function of the constants of motion of the Hamiltonian dynamics
(energy $E=H$, angular momentum ${L}=\gamma \sum_{i}r_{i}^{2}$ if the
domain has rotational symmetry and impulse ${\bf P}=\gamma\sum_{i}
{\bf r}_{i}$ if the domain has translational symmetries) is a
stationary solution of Eq. (\ref{b3}). For brevity, we shall only
consider the conservation of energy (the case of an infinite domain
with conservation of angular momentum is treated in \cite{lp}). The
basic postulate of statistical mechanics states that, at equilibrium,
all microscopic configurations that are accessible (i.e. that have the
correct value of energy) are {\it equiprobable}.  There is no
guarantee that the dynamics will lead the system to that ``uniform''
state because we could imagine that some regions of the
$2N$-dimensional phase space could be more probable than
others. However, if we accept this postulate, the equilibrium $N$-body
distribution is given by the microcanonical distribution
\begin{equation}
\label{b6} P_{N}({\bf r}_{1},...,{\bf r}_{N})={1\over
g(E)} \delta\lbrack E-H({\bf r}_{1},...,{\bf r}_{N})\rbrack.
\end{equation}
Using the normalization condition $\int P_{N}\prod_{i}d{\bf
r}_{i}=1$, we deduce that the density of states with energy $E$ is
given by
\begin{equation}
\label{b7} g(E)=\int  \delta\lbrack E-H({\bf r}_{1},...,{\bf r}_{N})\rbrack\prod_{i}d{\bf
r}_{i}.
\end{equation}
The microcanonical entropy of the system is defined by $S(E)=\ln
g(E)$ and the microcanonical temperature by $1/T(E)=\partial
S/\partial E$ (we take the Boltzmann constant $k_B=1$).  We
introduce the reduced probability distributions
\begin{equation}
\label{b8} P_{j}({\bf r}_{1},...,{\bf r}_{j})=\int P_{N}({\bf
r}_{1},...,{\bf r}_{N})d{\bf r}_{j+1}...d{\bf r}_{N}.
\end{equation}
For identical particles, the smooth (average) vorticity field  is
related to the one-body distribution function by
\begin{equation}
\label{b9} \omega({\bf r})=\langle \sum_{i=1}^{N}\gamma\delta({\bf
r}-{\bf r}_{i})\rangle =N\gamma P_{1}({\bf r}).
\end{equation}
Note that the vorticity field is proportional to the density of
point vortices: $\omega({\bf r})=\gamma n({\bf r})$. The total
circulation is $\Gamma=\int\omega({\bf r}) d{\bf r}=N\gamma$ and the
average value of the energy is
\begin{eqnarray}
\label{b10} E=\langle H\rangle={1\over 2}N(N-1)\gamma^{2}\int
u(|{\bf r}-{\bf r}'|)P_{2}({\bf r},{\bf r}')d{\bf r}d{\bf r}'.
\end{eqnarray}
By differentiating the defining relation for $P_j$ and using Eq.
(\ref{b6}), we can obtain an equilibrium hierarchy of equations
for the reduced moments \cite{lp}:
\begin{equation}
\label{b11} {\partial P_{j}\over\partial {\bf r}_{1}}=-{1\over
g(E)}{\partial\over\partial E}\biggl\lbrack
g(E)P_{j}\biggr\rbrack\sum_{i=2}^{j}\gamma^{2}{\partial
u_{1,i}\over\partial {\bf r}_{1}}-(N-j)\gamma^{2}\int {\partial
u_{1,j+1}\over\partial {\bf r}_{1}}{1\over
g(E)}{\partial\over\partial E}\biggl\lbrack g(E)P_{j+1}\biggr\rbrack
d{\bf r}_{j+1}.
\end{equation}
This is the counterpart of the equilibrium hierarchy in plasma physics. It
is however more complex in the present situation because it has been
derived in the microcanonical ensemble. Since statistical ensembles
are generically inequivalent for systems with long-range
interactions, we must formulate the problem in the microcanonical
ensemble which is the fundamental one. We note that the terms
involving the derivative of the density of states with respect to
energy can be split in two parts according to
\begin{equation}
\label{b12}
{1\over g(E)}{\partial\over\partial E}\biggl\lbrack g(E)P_{j}\biggr\rbrack=\beta P_{j}+{\partial P_{j}\over\partial E}.
\end{equation}
The terms with the $E$ derivative would not have emerged if we had
started from the Gibbs canonical distribution \cite{mj}. The
equivalent hierarchy of equations for material particles in
interaction is given in \cite{hb1}.

\subsection{Thermodynamic limit and mean field approximation} \label{sec_tl}

Since systems with long-range interactions are generically spatially
inhomogeneous, the usual thermodynamic limit $N,V\rightarrow +\infty$
with $N/V$ fixed is clearly irrelevant. We define the proper
thermodynamic limit of the point vortex gas as $N\rightarrow +\infty$
in such a way that the dimensionless temperature $\eta={\beta
N\gamma^{2}}$ and the dimensionless energy $\epsilon={E/ (N^2
\gamma^{2})}$ are fixed. It is convenient to rescale the parameters such
that $\gamma\sim 1/N$, $E\sim 1$, $\beta\sim N$ and $V\sim 1$. Then,
the total circulation $\Gamma=N\gamma$ remains of order unity. We note
that the ratio of $\partial P_{j}/\partial E$ on $\beta P_{j}$ is of
order $1/(E\beta)=1/(\epsilon\eta N)$. Therefore, in the thermodynamic
limit $N\rightarrow +\infty$ with $\epsilon$, $\eta$ fixed, the second
term in Eq. (\ref{b12}) is always negligible with respect to the
first. Using this simplification in the second equation of the
equilibrium hierarchy, we get
\begin{eqnarray}
\label{tl1} {\partial P_{1}\over\partial {\bf r}_{1}}({\bf r}_{1})=-\beta (N-1) \gamma^{2}\int P_{2}({\bf r}_{1},{\bf r}_{2}){\partial u_{12}\over\partial {\bf r}_{1}}d{\bf r}_{2}
-(N-1) \gamma^{2}\int {\partial u_{12}\over\partial {\bf r}_{1}}{\partial P_{2}\over\partial E}d{\bf r}_{2},
\end{eqnarray}
\begin{eqnarray}
\label{tl2} {\partial P_{2}\over\partial {\bf r}_{1}}({\bf
r}_{1},{\bf r}_{2})=-\beta \gamma^{2}P_{2}({\bf r}_{1},{\bf
r}_{2}){\partial u_{12}\over\partial {\bf r}_{1}}-\beta (N-2)
\gamma^{2}\int P_{3}({\bf r}_{1},{\bf r}_{2},{\bf r}_{3}) {\partial
u_{13}\over\partial {\bf r}_{1}} d{\bf r}_{3}.
\end{eqnarray}
We now decompose the two- and three-body distribution functions in
the suggestive form
\begin{equation}
\label{tl3}
P_{2}({\bf r}_{1},{\bf r}_{2})=P_{1}({\bf r}_{1})P_{1}({\bf r}_{2})+P_{2}'({\bf r}_{1},{\bf r}_{2}),
\end{equation}
\begin{eqnarray}
\label{tl4}
P_{3}({\bf r}_{1},{\bf r }_{2},{\bf r }_{3})=P_{1}({\bf r }_{1})P_{1}({\bf r }_{2})P_{1}({\bf r }_{3})+P_{2}'({\bf r }_{1},{\bf r }_{2})P_{1}({\bf r }_{3})\nonumber\\
+P_{2}'({\bf r }_{1},{\bf r }_{3})P_{1}({\bf r }_{2})+P_{2}'({\bf r }_{2},{\bf r }_{3})P_{1}({\bf r }_{1})+P_{3}'({\bf r }_{1},{\bf r }_{2},{\bf r }_{3}).
\end{eqnarray}
This decomposition is the counterpart of the first terms of the Mayer
expansion in plasma physics. The $P'_j$ are called the cumulants or
the correlation functions. Inserting these decompositions in
Eqs. (\ref{tl1})-(\ref{tl2}), we find after simplification that the
first two equations of the equilibrium hierarchy can be written
\begin{eqnarray}
\label{tl5} {\partial P_{1}\over\partial {\bf r}_{1}}({\bf r}_{1})=-\beta (N-1) \gamma^{2} P_{1}({\bf r}_{1})\int P_{1}({\bf r}_{2}){\partial u_{12}\over\partial {\bf r}_{1}}d{\bf r}_{2}\nonumber\\
-\beta (N-1) \gamma^{2}\int P_{2}'({\bf r}_{1},{\bf r}_{2}){\partial u_{12}\over\partial {\bf r}_{1}}d{\bf r}_{2}-(N-1) \gamma^{2}\int {\partial u_{12}\over\partial {\bf r}_{1}}{\partial P_{2}\over\partial E}({\bf r}_{1},{\bf r}_{2})d{\bf r}_{2},
\end{eqnarray}
\begin{eqnarray}
\label{tl6}
{\partial P_{2}'\over\partial {\bf r}_{1}}({\bf r}_{1},{\bf r}_{2})-(N-1)\gamma^{2}
P_{1}({\bf r}_{2})\int {\partial u_{13}\over\partial {\bf r}_{1}}{\partial P_{2}\over\partial E}({\bf r}_{1},
{\bf r}_{3})d {\bf r}_{3}=\nonumber\\
-\beta \gamma^{2} P_{1}({\bf r}_{1})P_{1}({\bf r}_{2}){\partial u_{12}\over\partial {\bf r}_{1}}
-\beta \gamma^{2} P_{2}'({\bf r}_{1},{\bf r}_{2}){\partial u_{12}\over\partial {\bf r}_{1}}\nonumber\\
+\beta \gamma^{2}P_{1}({\bf r}_{1})P_{1}({\bf r}_{2})
\int \frac{\partial u_{13}}{\partial {\bf r}_{1}}P_{1}({\bf r}_{3})
d{\bf r}_{3}-\beta (N-2) \gamma^{2}P_{2}'({\bf r}_{1},{\bf r}_{2})\int
P_{1}({\bf r}_{3}){\partial u_{13}\over\partial {\bf r}_{1}}d {\bf r}_{3}\nonumber\\
+\beta \gamma^{2}P_{1}({\bf r}_{2})\int \frac{\partial u_{13}}{\partial {\bf r}_{1}}P_{2}'({\bf r}_{1},{\bf r}_{3})d{\bf r}_{3}-\beta (N-2) \gamma^{2} P_{1}({\bf r}_{1})\int  P_{2}'({\bf r}_{2},{\bf r}_{3}){\partial u_{13}\over\partial {\bf r}_{1}}d {\bf r}_{3}\nonumber\\
-\beta (N-2) \gamma^{2}\int  P_{3}'({\bf r}_{1},{\bf r}_{2},{\bf
r}_{3}){\partial u_{13}\over\partial {\bf r}_{1}}d {\bf r}_{3},
\end{eqnarray}
where we have used Eq. (\ref{tl5}) to simplify some terms in
Eq. (\ref{tl6}). In the thermodynamic limit defined previously, it can
be shown that the correlation functions $P_{n}'$ are of order
$N^{-(n-1)}$ \cite{lp}.  Here, we shall just establish this result for
the two-body distribution function $P_{2}'$ assuming that it holds at
higher orders.  We thus neglect the term $P_{3}'$, of order $N^{-2}$,
in Eq. (\ref{tl6}). This is the counterpart of the Kirkwood
approximation in plasma physics. Then, considering the scaling of the
terms in Eq. (\ref{tl6}), we see that $P_{1}\sim 1$ and $P_{2}'\sim
\beta \gamma^{2}=\eta/N=O(1/N)$.  Therefore,
\begin{equation}
\label{tl7}
P_{2}({\bf r}_{1},{\bf r}_{2})=P_{1}({\bf r}_{1})P_{1}({\bf r}_{2})+O(1/N),
\end{equation}
so that the mean field approximation $P_{2}({\bf r}_{1},{\bf
r}_{2})\simeq P_{1}({\bf r}_{1})P_{1}({\bf r}_{2})$ is exact for
$N\rightarrow +\infty$. The coupling constant $\beta \gamma^{2}\sim
1/N$, scaling like the inverse of the point vortex number, plays a
role similar to the ``plasma parameter'' in plasma physics.

\subsection{The mean field equilibrium distribution} \label{sec_mfeq}

Taking the limit $N\rightarrow +\infty$ and using Eq. (\ref{tl7}),
the first equation (\ref{tl5}) of the equilibrium hierarchy
becomes
\begin{equation}
\label{mfeq1} \nabla\omega({\bf r})=-\beta \gamma\omega({\bf
r})\nabla\int\omega({\bf r}')u(|{\bf r}-{\bf r}'|)d{\bf r}',
\end{equation}
where $\omega({\bf r})=N \gamma P_{1}({\bf r})$ is the smooth
vorticity field. After integration, this can be written in the form
of the Boltzmann distribution
\begin{equation}
\label{mfeq2} \omega({\bf r})=Ae^{-\beta \gamma\psi({\bf r})},
\end{equation}
where
\begin{equation}
\label{mfeq3} \psi({\bf r})=\int \omega({\bf r}')u(|{\bf r}-{\bf
r}'|)d{\bf r}',
\end{equation}
is the stream function produced by the smooth distribution of
point vortices. Therefore, the equilibrium density profile of the
point vortices is determined by an {integrodifferential} equation. For
the usual potential of interaction, satisfying $\Delta u=-\delta$,
we find that the equilibrium vorticity profile is determined by the
Boltzmann-Poisson equation
\begin{equation}
\label{mfeq4} -\Delta\psi=Ae^{-\beta \gamma\psi({\bf r})}.
\end{equation}
These results can also be obtained by maximizing the Boltzmann entropy
at fixed circulation and energy in order to obtain the {\it most
probable} distribution of point vortices at statistical equilibrium
\cite{jm}. These results can be generalized so as to take into account
the conservation of the angular momentum. In that case, the stream
function $\psi$ in the Boltzmann distribution is replaced by the
relative stream function $\psi'=\psi+(1/2)\Omega_L r^2$
\cite{chavlemou} where $\Omega_{L}$ is a Lagrange multiplier associated with the conservation of the angular momentum (the conservation of the linear impulse can be dealt with similarly \cite{jfm2}).

\section{Kinetic equation from a BBGKY-like hierarchy}
\label{sec_bbgky}

\subsection{The BBGKY-like hierarchy}
\label{sec_h}

We now address the out-of-equilibrium problem by using a methodology
similar to the previous one. Our aim is to derive a kinetic equation
for the evolution of the smooth vorticity profile $\omega({\bf r},t)$
of the point vortex gas and to see whether or not it converges to the
statistical equilibrium state (\ref{mfeq2}).  Integrating the
Liouville equation (\ref{b3}) on ${\bf r}_{j+1}$,...,${\bf r}_{N}$, it
is simple to construct a hierarchy of equations for the reduced
distributions. It has the form
\begin{eqnarray}
\frac{\partial P_j}{\partial t}+\sum_{i=1}^{j}\sum_{k=1,k\neq
i}^{j}{\bf V}(k\rightarrow i)\frac{\partial P_j}{\partial {\bf
r}_i}+(N-j)\sum_{i=1}^{j}\int {\bf V}(j+1\rightarrow
i)\frac{\partial P_{j+1}}{\partial {\bf r}_i}d{\bf r}_{j+1}=0.
 \label{h1}
\end{eqnarray}
This is the counterpart of the BBGKY hierarchy in plasma physics.
The first two equations of this hierarchy are
\begin{eqnarray}
\frac{\partial P_1}{\partial t}+(N-1)\frac{\partial}{\partial {\bf
r}_1} \int {\bf V}(2\rightarrow 1)P_2({\bf r}_1,{\bf r}_2) d{\bf
r}_{2}=0,
 \label{h2}
\end{eqnarray}
\begin{eqnarray}
\frac{\partial P_2}{\partial t}+{\bf V}(2\rightarrow 1)
\frac{\partial P_2}{\partial {\bf
r}_1}+(N-2)\frac{\partial}{\partial {\bf r}_1} \int {\bf
V}(3\rightarrow 1)P_3({\bf r}_1,{\bf r}_2,{\bf r}_3) d{\bf
r}_{3}+(1\leftrightarrow 2)=0.
 \label{h3}
\end{eqnarray}
For brevity, we have not written the variable $t$ in the distribution
functions. Inserting the decomposition (\ref{tl3}) in
Eq. (\ref{h2}), we first obtain
\begin{eqnarray}
\frac{\partial P_1}{\partial t}+(N-1) \frac{\partial P_1}{\partial
{\bf r}_1} \int {\bf V}(2\rightarrow 1)P_1({\bf r}_2) d{\bf
r}_{2}+(N-1)\frac{\partial}{\partial {\bf r}_1} \int {\bf
V}(2\rightarrow 1)P_2'({\bf r}_1,{\bf r}_2) d{\bf r}_{2}=0.
 \label{h4}
\end{eqnarray}
Next, substituting the decomposition (\ref{tl3}) and (\ref{tl4}) in
Eq. (\ref{h3}) and using (\ref{h4}) to simplify some terms, we obtain
\begin{eqnarray}
\frac{\partial P_2'}{\partial t}+{\bf V}(2\rightarrow 1)
\frac{\partial P_2'}{\partial {\bf r}_1}+{\bf
V}(2\rightarrow 1)P_1({\bf r}_2) \frac{\partial P_1}{\partial {\bf r}_1}({\bf
r}_1)\nonumber\\
-P_{1}({\bf r}_{2})\frac{\partial}{\partial{\bf r}_{1}}\int {\bf V}(3\rightarrow 1)P_{1}({\bf r}_{1})P_{1}({\bf r}_{3}) d{\bf r}_{3}
\nonumber\\
-\frac{\partial}{\partial{\bf r}_{1}}\int {\bf V}(3\rightarrow 1)P_{2}'({\bf r}_{1},{\bf r}_{3})P_{1}({\bf r}_{2}) d{\bf r}_{3}\nonumber\\
+(N-2)\frac{\partial}{\partial {\bf r}_1} \int {\bf V}(3\rightarrow
1)P_2'({\bf r}_1,{\bf r}_2)P_1({\bf r}_3) d{\bf
r}_{3}\nonumber\\
+(N-2)\frac{\partial}{\partial {\bf r}_1} \int {\bf V}(3\rightarrow
1)P_2'({\bf r}_2,{\bf r}_3)P_1({\bf r}_1) d{\bf
r}_{3}\nonumber\\
+(N-2)\frac{\partial}{\partial {\bf r}_1} \int {\bf V}(3\rightarrow
1)P_3'({\bf r}_1,{\bf r}_2,{\bf r}_3)d{\bf
r}_{3}+(1\leftrightarrow 2)=0.
 \label{h5}
\end{eqnarray}
The equations (\ref{h4}) and (\ref{h5}) are exact for all $N$ but the
hierarchy is not closed. We shall now consider the thermodynamic limit
defined in Sec. \ref{sec_tl}. Based on the scaling of the terms in
each equation of the hierarchy, we argue that there exists solutions
of the whole BBGKY-like hierarchy such that the correlation functions
$P_{j}'$ scale like $1/N^{j-1}$ at any time. This implicitly assumes
that the initial condition has no correlation, or that the initial
correlations respect this scaling (if there are strong initial
correlations, like vortex pairs, the kinetic theory will be different
from the one developed in the sequel). Recalling that $P_1\sim 1$,
$P_2'\sim 1/N$ and $|{\bf V}(i\rightarrow j)|\sim\gamma\sim 1/N$, we
obtain at order $1/N$:
\begin{eqnarray}
\frac{\partial P_1}{\partial t}+(N-1) \frac{\partial P_1}{\partial
{\bf r}_1} \int {\bf V}(2\rightarrow 1)P_1({\bf r}_2) d{\bf
r}_{2}+N\frac{\partial}{\partial {\bf r}_1} \int {\bf
V}(2\rightarrow 1)P_2'({\bf r}_1,{\bf r}_2) d{\bf r}_{2}=0,
 \label{h6}
\end{eqnarray}
\begin{eqnarray}
\label{h7} {\partial P_{2}'\over\partial t}+ \left \lbrack {\bf V}(2\rightarrow 1)-\int {\bf V}(3\rightarrow 1)P_{1}({\bf r}_{3})d{\bf r}_{3}\right \rbrack P_{1}({\bf r}_{2}){\partial P_{1}\over\partial {\bf r}_{1}}({\bf r}_{1})
\nonumber\\
+N{\partial P_{2}'\over\partial {\bf r}_{1}}\int {\bf V}(3\rightarrow 1)P_{1}({\bf r}_{3})d{\bf r}_{3} +N{\partial\over\partial {\bf r}_{1}}\int {\bf V}(3\rightarrow 1)P_{2}'({\bf r}_{2},{\bf r}_{3})P_{1}({\bf r}_{1})d{\bf r}_{3}+(1\leftrightarrow 2)=0.
\end{eqnarray}
The three-body correlation function can be neglected. If we introduce the notations $\omega=N\gamma P_{1}$ (smooth vorticity field) and $g=N^{2}P_{2}'$ (two-body correlation function), these equations can be rewritten
\begin{eqnarray}
{\partial \omega_{1}\over\partial t}+\frac{N-1}{N}\langle {\bf V}\rangle_{1} {\partial \omega_{1}\over \partial {\bf r}_{1}}=-\gamma {\partial \over\partial {\bf r}_{1}}\int {\bf V}(2\rightarrow 1)g({\bf r}_{1},{\bf r}_{2})d{\bf r}_{2},
\label{h8}
\end{eqnarray}
\begin{eqnarray}
\label{h9} {\partial g\over\partial t}+\langle {\bf V}\rangle_{1} {\partial g\over\partial {\bf r}_{1}}+\frac{1}{\gamma^{2}}
 {\bf {\cal V}}(2\rightarrow 1)\omega_{2}  {\partial \omega_{1}\over\partial {\bf r}_{1}}\nonumber\\
+\frac{\partial}{\partial {\bf r}_{1}} \int {\bf V}(3\rightarrow
1)g({\bf r}_{2},{\bf r}_{3},t) \frac{\omega_{1}}{\gamma}d{\bf
r}_{3}+(1\leftrightarrow 2)=0.
\end{eqnarray}
For brevity, we have used the abbreviations  $\omega_{1}=\omega({\bf r}_{1},t)$
and $\omega_{2}=\omega({\bf r}_{2},t)$. We have also introduced the
mean velocity in ${\bf r}_1$ created by all the vortices
\begin{eqnarray}
\label{h10}
\langle {\bf V}\rangle_{1} =\int {\bf V}(2\rightarrow 1)\frac{\omega_2}{\gamma}d{\bf r}_{2},
\end{eqnarray}
and the fluctuating velocity created by point vortex $2$ on point
vortex $1$:
\begin{eqnarray}
\label{h11}
{\bf {\cal V}}(2\rightarrow 1)={\bf V}(2\rightarrow 1)-\frac{1}{N}\langle {\bf V}\rangle_{1}.
\end{eqnarray}
These equations (\ref{h8})-(\ref{h9}) are exact at order
$O(1/N)$. They form therefore the right basis to develop a kinetic
theory.

{\it (i) Collisionless regime:} If we consider the limit $N\rightarrow
+\infty$ (for a fixed time $t$), noting that $P_{2}'=O(1/N)\rightarrow
0$, we find that the smooth vorticity field $\omega({\bf r},t)$ of the
point vortex gas is solution of the 2D Euler equation
\begin{eqnarray}
\label{h12}
{\partial \omega\over\partial t}+\langle {\bf
V}\rangle\nabla\omega=0,\qquad \langle {\bf V}\rangle =-{\bf
z}\times\nabla\psi,\label{h4ff}
\end{eqnarray}
where the stream function $\psi({\bf r},t)$ is given by
Eq. (\ref{mfeq3}) with $\omega({\bf r},t)$ instead of $\omega({\bf
r})$. The 2D Euler equation is valid when the correlations between
point vortices can be neglected, i.e. $P_{2}({\bf r}_1,{\bf r
}_2,t)=P_{1}({\bf r}_1,t)P_{2}({\bf r}_2,t)$, which is the case for
$N\rightarrow +\infty$. The Euler equation describes the {\it
collisionless evolution} of the point vortex gas up to a time of order
$N t_D$ (where $t_D$ is the dynamical time) at least. In practice,
$N\ge 1000$ so that the domain of validity of the 2D Euler equation is
huge. The Euler equation is the counterpart of the Vlasov equation in
plasma physics and stellar dynamics.  It can undergo a process of
violent relaxation towards a Quasi Stationary State
\cite{lb,miller,rs,csr} as discussed in 
Secs. \ref{sec_v} and \ref{sec_ivr}. 

{\it (ii) Collisional regime:} If we want to describe the collisional
evolution of the point vortex gas, we need to consider finite $N$
effects. Equations (\ref{h8})-(\ref{h9}) describe the evolution of the
system on a timescale of order $N t_D$. The equation for the evolution
of the smooth vorticity field is of the form
\begin{eqnarray}
{\partial \omega\over\partial t}+\frac{N-1}{N}\langle {\bf
V}\rangle\nabla\omega=C_{N}[\omega],\label{h13}
\end{eqnarray}
where $C_{N}$ is a ``collision'' term analogous to the one arising in
the Boltzmann equation. In the present context, there are not real
collisions between point vortices. The term on the right hand side of
Eq. (\ref{h13}) is due to the development of correlations between
vortices as time goes on. It is related to the two-body correlation
function $g({\bf r}_{1},{\bf r}_{2},t)$ which is determined in terms
of the vorticity by Eq. (\ref{h9}).  Our aim is to obtain an
expression for the collision term $C_{N}[\omega]$ at the order
$1/N$. The difficulty with Eq. (\ref{h9}) for the two-body correlation
function is that it is an integrodifferential equation. The second
term is an advective term, the third term is the {\it source} of the
correlation and the fourth term takes into account {\it collective
effects}. In this paper, we shall neglect the contribution of the
integral in Eq.  (\ref{h9}). Then, we get the coupled system
\begin{eqnarray}
{\partial \omega_{1}\over\partial t}+\frac{N-1}{N}\langle {\bf
V}\rangle_{1} {\partial \omega_1\over \partial {\bf r}_{1}}=-\gamma
{\partial \over\partial {\bf r}_{1}}\int {\bf V}(2\rightarrow
1)g({\bf r}_{1},{\bf r}_{2})d{\bf r}_{2}, \label{h14}
\end{eqnarray}
\begin{eqnarray}
\label{h15} {\partial g\over\partial t}+\left \lbrack \langle {\bf
V}\rangle_{1} {\partial\over\partial {\bf r}_{1}}+\langle {\bf
V}\rangle_{2} {\partial\over\partial {\bf r}_{2}}\right  \rbrack g
+\left \lbrack {\bf {\cal V}}(2\rightarrow 1) {\partial\over\partial
{\bf r}_{1}}+{\bf {\cal V}}(1\rightarrow 2) {\partial\over\partial
{\bf r}_{2}}\right \rbrack
\frac{\omega_{1}}{\gamma}\frac{\omega_{2}}{\gamma}=0.
\end{eqnarray}
The integral that we have neglected contains ``collective effects''
that are taken into account in the approach of Dubin \& O'Neil
\cite{dn}.  However, their study is restricted to axisymmetric flows
and makes a Markovian approximation. These assumptions are necessary
to use Laplace-Fourier transforms in order to solve the
integro-differential equation (\ref{h9}). Alternatively, if we neglect
collective effects,  we can obtain a general kinetic
equation in a closed form (\ref{gen6}) that is valid for flows that are not
necessarily axisymmetric and that can take into account memory
effects. This equation has interest in its own right (despite its
limitations) because its structure bears a lot of physical
significance.  Before deriving this general equation, we shall first
consider the case of axisymmetric flows and obtain a simple explicit
kinetic equation valid for such flows when collective effects are
neglected.

\subsection{The Markovian axisymmetric equation}
\label{sec_ma}

For an axisymmetric flow, the vorticity field and the two-point
correlation function can be written as $\omega=\omega(r,t)$ and
$g=g(r_1,r_2,\theta_1-\theta_2,t)$, and the mean velocity as
$\langle {\bf V}\rangle=\langle V\rangle_{\theta} (r,t){\bf e}_{\theta}$. On
the other hand, the projection of ${\bf V}(2\rightarrow 1)$ in the
direction of ${\bf r}_1$ is
\begin{eqnarray}
{V}_{r_1}(2\rightarrow 1)=\gamma \frac{1}{{r}_{1}}\frac{\partial
u_{12}}{\partial \theta_1},\label{ma1}
\end{eqnarray}
where $u_{12}=u(r_1,r_2,\theta_1-\theta_2)$ is symmetric in $r_1$ and
$r_2$ (see Appendix \ref{sec_pot}). In that case,
Eqs. (\ref{h14})-(\ref{h15}) become
\begin{eqnarray}
\frac{\partial \omega_1}{\partial t}=-\gamma^2
\frac{1}{r_1}\frac{\partial}{\partial {r}_1} \int_0^{+\infty}r_2
dr_2\int_{0}^{2\pi} \frac{\partial u}{\partial \phi}
g({r}_1,{r}_2,\phi,t)d\phi, \label{ma2}
\end{eqnarray}
\begin{eqnarray}
\frac{\partial g}{\partial t}+\left\lbrack
\Omega(r_1,t)-\Omega(r_2,t)\right\rbrack  \frac{\partial g}
{\partial \phi}=-\frac{\partial u}{\partial \phi} \left
(\frac{1}{r_1}\frac{\partial}{\partial
{r_1}}-\frac{1}{r_2}\frac{\partial}{\partial r_2}\right
)\omega({r}_1,t)\frac{\omega}{\gamma}({r}_2,t),\label{ma3}
\end{eqnarray}
where we have set $\phi=\theta_1-\theta_2$ and where
$\Omega(r,t)=\langle V\rangle_{\theta}(r,t)/r$ is the angular
velocity of the mean flow. Taking the Fourier transform of Eq. (\ref{ma3})
with respect to $\phi$ and introducing the notations
$\partial=(1/r_1){\partial}/\partial
{r}_1-(1/r_2){\partial}/{\partial {r}_2}$,
$\omega_1=\omega({r}_1,t)$, $\omega_2=\omega({r}_2,t)$ and
$\Delta\Omega=\Omega(r_1,t)-\Omega(r_2,t)$, we obtain
\begin{eqnarray}
\frac{d \hat{g}_m}{d t}+i m \Delta\Omega\hat
g_m=-\frac{i}{\gamma}m\hat{u}_m \partial \omega_1
\omega_2.\label{ma4}
\end{eqnarray}
The Fourier transform of the potential of interaction $u$ is discussed
in Appendix \ref{sec_pot} where explicit examples are considered. In
terms of the Fourier transform of the correlation function, the
kinetic equation (\ref{ma2}) can be rewritten
\begin{eqnarray}
\frac{\partial \omega_1}{\partial t}=-2\pi\gamma^2
\frac{1}{r_1}\frac{\partial}{\partial {r}_1}\int_0^{+\infty}r_2 dr_2
\sum_m m  \hat{u}_m {\rm Im}\hat{g}_m(r_1,r_2,t). \label{ma5}
\end{eqnarray}
We shall assume that ${\rm Im}\hat g_m(r_1,r_2,t)$ relaxes on a
timescale which is much smaller than the timescale on which
$\omega(r,t)$ changes. This is the equivalent of the Bogoliubov
hypothesis in plasma physics.  If we ignore memory effects, we
can integrate the first order differential equation (\ref{ma4}) by
considering the last term as a constant. This yields
\begin{eqnarray}
\hat{g}_m({r}_1,{r}_2,t)=-\int_0^t d\tau \frac{i}{\gamma}m\hat{u}_m
e^{-i m\Delta\Omega\tau}\partial \omega_1\omega_2,\label{ma6}
\end{eqnarray}
where we have assumed that no correlation is present initially:
$g(t=0)=0$. Then, we can replace ${\rm Im}\hat{g}_m(r_1,r_2,t)$ in
Eq. (\ref{ma5}) by its value obtained for $t\rightarrow +\infty$, which
reads
\begin{eqnarray}
{\rm Im}\hat{g}_m({r}_1,{r}_2,+\infty)=
-\frac{\pi}{\gamma}m\hat{u}_m \delta(m\Delta\Omega)\partial
\omega_1\omega_2.\label{ma7}
\end{eqnarray}
Substituting this relation in Eq. (\ref{ma5}), we obtain the kinetic
equation
\begin{eqnarray}
\frac{\partial \omega_1}{\partial t}=2\pi^2 \gamma \frac{1}{r_1}
\frac{\partial}{\partial r_1} \int_0^{+\infty} r_2 dr_2
\chi(r_1,r_2) \delta(\Omega_1-\Omega_2) \left
(\frac{1}{r_1}\omega_2\frac{\partial \omega_1}{\partial
r_1}-\frac{1}{r_2}\omega_1\frac{\partial \omega_2}{\partial
 r_2}\right ), \label{ma8}
\end{eqnarray}
where we have defined
\begin{eqnarray}
\chi(r_1,r_2)=\sum_m {|m|}\hat{u}_m(r_{1},r_{2})^2. \label{ma9}
\end{eqnarray}
For the potential of interaction (\ref{pot10}), this function is given by Eq.
(\ref{pot13}) and we recover the kinetic equation obtained in \cite{kin}:
\begin{eqnarray}
\frac{\partial \omega_1}{\partial t}=-\frac{\gamma}{4} \frac{1}{r_1}
\frac{\partial}{\partial r_1} \int_0^{+\infty} r_2 dr_2
\ln\left\lbrack 1-\left (\frac{r_<}{r_>}\right
)^{2}\right\rbrack\delta(\Omega_1-\Omega_2) \left
(\frac{1}{r_1}\omega_2\frac{\partial \omega_1}{\partial
r_1}-\frac{1}{r_2}\omega_1\frac{\partial \omega_2}{\partial
 r_2}\right ). \label{ma10}
\end{eqnarray}
This equation, which ignores collective effects, is the vortex
analogue of the Landau equation in plasma physics. We can show
\cite{kin,chavlemou} that it conserves $\Gamma$, $E$, $L$, that it
satisfies an $H$-theorem ($\dot S\ge 0$) and that the Boltzmann
distribution (\ref{g2}) is a particular steady state, but not the only
one (see \cite{chavlemou} for more discussion). Collective effects can
be taken into account by keeping the contribution of the last integral
in Eq. (\ref{h9}). For axisymmetric flows, these terms could be
evaluated at the price of complicated calculations in the complex
plane similar to those performed in plasma physics to derived the
Lenard-Balescu equation from the BBGKY hierarchy. It would be
interesting to make this derivation although it will not be attempted
in the present paper. This would certainly lead to the kinetic
equation derived by Dubin \& O'Neil \cite{dn} from a quasilinear
theory of the Klimontovich equation. As we shall see, the
consideration of collective effects is equivalent to replacing the
bare potential of interaction by an ``effective potential''. The
resulting kinetic equation remains of the form of Eq. (\ref{ma8})
with a modified function $\chi_{P}(r_1,r_2)$. Therefore, as far as the
general {\it structure} of the kinetic equations is concerned, our
simple treatment is of interest. Furthermore, it can be generalized to
non axisymmetric flows as considered in the next section. Finally,
since the diffusion coefficient in Eq. (\ref{ma8}) does not diverge
(contrary to the 3D Landau equation in plasma physics), the
Lenard-Balescu treatment of collective effects is not necessary in our
case for a first analysis.

\subsection{The general non Markovian kinetic equation}
\label{sec_gen}

The above kinetic equations assume that the flow is axisymmetric and
rely on the assumption that the correlation function relaxes much more
rapidly than the vorticity field. The Markovian approximation is
expected to be a good approximation in the limit $N\rightarrow
+\infty$ that we consider since the vorticity profile changes only on
a timescale of order $N t_D$ (where $t_D$ is the dynamical time) or
even larger. However, for systems with long-range interactions, there
can be situations where the decorrelation time of the fluctuations is
very long so that the Markovian approximation may not be completely
justified (this is the case for self-gravitating systems).  For
comparison, and for sake of generality, it can be of interest to
derive non-markovian kinetic equations for point vortices. For an
{\it arbitrary} flow, Eq. (\ref{h15}) for the correlation function can be
written
\begin{eqnarray}
\label{gen3} {\partial g\over\partial t}+{\cal L} g=-\left \lbrack
{\bf {\cal V}}(2\rightarrow 1) {\partial\over\partial {\bf r}_{1}}
+{\bf {\cal V}}(1\rightarrow 2) {\partial\over\partial {\bf r}_{2}}
\right \rbrack \frac{\omega}{\gamma}({\bf
r}_{1},t)\frac{\omega}{\gamma}({\bf r}_{2},t),
\end{eqnarray}
where we have denoted the advective term by ${\cal L}$ (Liouvillian
operator). Solving formally this equation with the Green function
\begin{eqnarray}
G(t,t')={\rm exp}\left\lbrace -\int_{t'}^t {\cal
L}(\tau)d\tau\right\rbrace,\label{gen4}
\end{eqnarray}
we obtain
\begin{eqnarray}
g({\bf r}_1,{\bf r}_2,t)=-\int_0^t d\tau G(t,t-\tau) \left \lbrack {\bf {\cal V}}(2\rightarrow 1) {\partial\over\partial {\bf r}_{1}}+{\bf {\cal V}}(1\rightarrow 2) {\partial\over\partial {\bf r}_{2}}\right \rbrack\frac{\omega}{\gamma}({\bf
r}_1,t-\tau)\frac{\omega}{\gamma}({\bf r}_2,t-\tau).\nonumber\\
\label{gen5}
\end{eqnarray}
The Green function constructed with the smooth velocity field $\langle
{\bf V}\rangle$ means that, in order to evaluate the time integral in
Eq. (\ref{gen5}), we must move the coordinates ${\bf r}_i(t-\tau)$ of
the point vortices with the mean field flow $\langle {\bf V}\rangle
({\bf r},t)$, adopting a Lagrangian point of view. Thus, in evaluating
the integral, the coordinates ${\bf r}_{i}$ must be viewed as ${\bf
r}_{i}(t-\tau)$, where ${\bf r}_{i}(t-\tau)={\bf
r}_{i}(t)-\int_{0}^{\tau}ds\ \langle {\bf V}\rangle ({\bf
r}_{i}(t-s),t-s)ds$.  Substituting this result in Eq. (\ref{h14}), we
obtain
\begin{eqnarray}
\frac{\partial \omega_1}{\partial t}+\frac{N-1}{N}\langle {\bf
V}\rangle_{1} {\partial \omega\over \partial {\bf r}_{1}}
=\frac{\partial}{\partial {r}_1^{\mu}}\int_0^t d\tau \int d{\bf
r}_{2} {V}^{\mu}(2\rightarrow
1,t)G(t,t-\tau)\nonumber\\
\times  \left \lbrack {{\cal V}}^{\nu}(2\rightarrow 1)
{\partial\over\partial { r}_{1}^{\nu}}+{{\cal V}}^{\nu}(1\rightarrow
2) {\partial\over\partial {r}_{2}^{\nu}}\right
\rbrack\omega({\bf
r}_1,t-\tau)\frac{\omega}{\gamma}({\bf r}_2,t-\tau). \label{gen6}
\end{eqnarray}
This returns the general kinetic equation obtained by Chavanis \cite{kin} with
the projection operator formalism (note that we can replace
${V}^{\mu}(2\rightarrow 1,t)$ by ${\cal V}^{\mu}(2\rightarrow 1,t)$ in
the first term of the r.h.s. of the equation since the fluctuations
vanish in average). It slightly differs from the equation obtained in
\cite{kin} by a term $(N-1)/N$ in the l.h.s. This new derivation of
the kinetic equation (\ref{gen6}) from a systematic expansion of the
solutions of the BBGKY hierarchy in powers of $1/N$ is valuable
because the formalism is much simpler than the projection operator
formalism and clearly shows which terms have been neglected in the
derivation. It also clearly shows that {\it the kinetic equation
(\ref{gen6}) is valid at order $1/N$ so that it describes the system
on a timescale of order $Nt_D$.} In
\cite{houches,chavlemou}, we had obtained this estimate a posteriori.

\subsection{Summary of the different kinetic equations}\label{sec_s}

Let us briefly summarize the different kinetic equations that appeared
in our analysis. When collective effects are ignored, the kinetic
equation describing the evolution of the system as a whole at order
$1/N$ is
\begin{eqnarray}
\frac{\partial \omega}{\partial t}+\frac{N-1}{N}\langle {\bf
V}\rangle \nabla\omega=\frac{\partial}{\partial
r^{\mu}}\int_{0}^{t}d\tau\int d{\bf r}_{1}{V}^{\mu}(1\rightarrow 0)
G(t,t-\tau)\nonumber\\
\times \biggl\lbrace {\cal V}^{\nu}(1\rightarrow
0)\frac{\partial}{\partial r^{\nu}} + {\cal V}^{\nu}(0\rightarrow
1)\frac{\partial}{\partial r_{1}^{\nu}}\biggr\rbrace
\omega({\bf r},t-\tau)\frac{\omega}{\gamma}({\bf r}_{1},t-\tau).
\label{s1}
\end{eqnarray}
If we make a Markov approximation and extend the time integral to infinity, we obtain
\begin{eqnarray}
\frac{\partial \omega}{\partial t}+\frac{N-1}{N}\langle {\bf
V}\rangle \nabla\omega=\frac{\partial}{\partial
r^{\mu}}\int_{0}^{+\infty}d\tau\int d{\bf r}_{1}{V}^{\mu}(1\rightarrow 0)
G(t,t-\tau)\nonumber\\
\times \biggl\lbrace {\cal V}^{\nu}(1\rightarrow
0)\frac{\partial}{\partial r^{\nu}} + {\cal V}^{\nu}(0\rightarrow
1)\frac{\partial}{\partial r_{1}^{\nu}}\biggr\rbrace
\omega({\bf r},t)\frac{\omega}{\gamma}({\bf r}_{1},t).
\label{s1b}
\end{eqnarray}
As we have indicated, the Markov approximation is justified for
$N\rightarrow +\infty$ because the timescale $Nt_{D}$ on which
$\omega$ changes is long compared to the timescale $\tau_{corr}$ for
which the integrand in Eq. (\ref{s1b}) has significant support. We do
not assume that the decorrelation time is extremely short so that, in
the time integral, the vorticity and the vorticity gradient must be
evaluated at ${\bf r}(t-\tau)$ and ${\bf r}_{1}(t-\tau)$ where now
${\bf r}_{i}(t-\tau)={\bf r}_{i}(t)-\int_{0}^{\tau}ds\ \langle {\bf
V}\rangle ({\bf r}_{i}(t-s),t)ds$. On the other hand, for an
axisymmetric evolution, using the relation (\ref{pot5}) and
$r_{i}(t-\tau)=r_{i}(t)$ and
$\theta_{i}(t-\tau)=\theta_{i}(t)-\Omega(r_{i}(t),t)\tau$,
Eq. (\ref{s1}) takes the form
\begin{eqnarray}
\frac{\partial \omega}{\partial
t}=\frac{1}{r}\frac{\partial}{\partial
r}r\int_{0}^{t}d\tau\int_{0}^{2\pi}d\theta_{1}\int_{0}^{+\infty} r
r_{1} d{r}_{1}{V}_{r}(1\rightarrow 0,t)\nonumber\\
\times {V}_{r}(1\rightarrow 0,t-\tau)\biggl
(\frac{1}{r}\frac{\partial}{\partial r}
-\frac{1}{r_{1}}\frac{\partial}{\partial r_{1}}\biggr )
{\omega}(r,t-\tau)\frac{\omega}{\gamma}({r}_{1},t-\tau). \label{s2}
\end{eqnarray}
The integral on $\theta_{1}$ can be performed using Eq.
(\ref{d10}), and we get
\begin{eqnarray}
\frac{\partial \omega}{\partial
t}=2\pi\gamma \frac{1}{r}\frac{\partial}{\partial
r}\int_{0}^{t}d\tau\int_{0}^{+\infty} 
r_{1} d{r}_{1}\sum_{m}m^{2}\hat{u}_{m}^{2}(r,r_{1})\cos(m\Delta\Omega\tau)\nonumber\\
\times \biggl
(\frac{1}{r}\frac{\partial}{\partial r}
-\frac{1}{r_{1}}\frac{\partial}{\partial r_{1}}\biggr )
{\omega}(r,t-\tau){\omega}({r}_{1},t-\tau). \label{s2nbt}
\end{eqnarray}
This equation can also be obtained from the approach of Sec. \ref{sec_ma} by
keeping memory effects in Eq. (\ref{ma6}).  If we make a Markovian
approximation ${\omega}({r}_{1},t-\tau)\simeq {\omega}({r}_{1},t)$ and
${\omega}({r},t-\tau)\simeq {\omega}({r},t)$, and extend the time
integration to $+\infty$ in Eq. (\ref{s2}), we obtain
\begin{eqnarray}
\frac{\partial \omega}{\partial
t}=\frac{1}{r}\frac{\partial}{\partial
r}r\int_{0}^{+\infty}d\tau\int_{0}^{2\pi}d\theta_{1}\int_{0}^{+\infty}
r r_{1} d{r}_{1}{V}_{r}(1\rightarrow 0,t)\nonumber\\
\times {V}_{r}(1\rightarrow 0,t-\tau)\biggl
(\frac{1}{r}\frac{\partial}{\partial r}
-\frac{1}{r_{1}}\frac{\partial}{\partial r_{1}}\biggr )
{\omega}(r,t)\frac{\omega}{\gamma}({r}_{1},t). \label{s3}
\end{eqnarray}
The integral on $\tau$ and $\theta_{1}$ can be performed using Eq.
(\ref{d10}), see \cite{kin,chavlemou} for details, and we get the
kinetic equation (\ref{ma8}). If we make the approximation
${\omega}({r}_{1},t-\tau)\simeq {\omega}({r}_{1},t)$ and
${\omega}({r},t-\tau)\simeq {\omega}({r},t)$ but keep the time
integration from $0$ to $t$, we obtain the equation derived in
\cite{chavlemou} incorporating a function $M(t)$ which regularizes
the delta function occuring in Eq. (\ref{ma8}). Finally, in Appendix
\ref{sec_heur}, we propose a simple heuristic kinetic equation that
may be of interest.

\subsection{Discussion}
\label{sec_disc}

These kinetic equations possess a lot of interesting properties. Let
us first consider the Markovian axisymmetric equation (\ref{ma8}).
The collisional evolution of point vortices is truly due to long range
interactions because the current in $r$ is caused by ``distant
collisions'' with vortices located in $r_1\neq r$ that can be far
away. This is different from plasma physics and stellar dynamics where
the collisions are assumed to be {\it local} \cite{bt}. Therefore, the
current occurs only in velocity space and is due to ``collisions''
involving particles at the same location but having different
velocities $v_1\neq v$ (recall that the position $r$ in the point
vortex system plays the same role as the velocity $v$ in the plasma
system). Therefore, in the case of stellar systems and plasmas, the
collisional term is determined as if the system were spatially
homogeneous. For these systems, long-range interactions manifest
themselves only as mean field effects in the advective term (Vlasov)
of the kinetic equation (see Eq. (44) of \cite{hb2}). By contrast, the
point vortex gas is the first system where collisions involve distant
particles. The collisional evolution in $r$ is due to a condition of
resonance $\Omega(r_1,t)=\Omega(r,t)$ with point vortices in $r_1\neq
r$ that have the same angular velocity.  Clearly, this condition can
be satisfied only when the profile of angular velocity is
non-monotonic \cite{dn,kin}. Therefore, the evolution stops when the
profile of angular velocity becomes monotonic (so that there is no
resonance) even if the system has not reached statistical
equilibrium. In that case, the system settles on a Quasi Stationary
State (QSS) that is {\it not} the Boltzmann distribution (\ref{mfeq2})
predicted by statistical mechanics \cite{chavlemou}. On the timescale
$Nt_D$ on which the kinetic theory is valid, the collisions tend to
create a monotonic profile of angular velocity. Since the entropy
increases monotonically, the vorticity profile {\it tends} to approach the
Boltzmann distribution but does not attain it in general because of
the absence of resonances.  The Boltzmann distribution may be reached
on longer timescales, larger than $Nt_D$. To describe this regime, we
need to determine terms of order $N^{-2}$ or smaller in the expansion
of the solutions of the BBGKY hierarchy for $N\rightarrow
+\infty$. This implies in particular the determination of the
three-body correlation function, which is a formidable task. It is
interesting to note that the markovian axisymmetric kinetic equation
(\ref{ma8}) conserves all the integral constraints of the point vortex
dynamics (circulation, energy, angular momentum) and satisfies an
H-theorem for the Boltzmann entropy, so that the entropy is
non-decreasing $\dot S\ge 0$
\cite{chavlemou}. However, as we have indicated previously, 
this kinetic equation does not in general converge towards the
Boltzmann distribution.  This is because this kinetic equation admits
an infinite number of stationary solutions among which the Boltzmann
distribution is just a particular case (see \cite{chavlemou} for a
detailed discussion). This is at variance with the Landau and
Lenard-Balescu equations which always converge towards the Boltzmann
distribution
\cite{hb2}. In these equations, the collisional evolution of the
system is also due to a condition of resonance ${\bf k}\cdot {\bf
v}_1={\bf k}\cdot {\bf v}$ (see Eq. (49) of \cite{hb2}) but the
Boltzmann distribution is the only steady state of these kinetic
equations. As noted in \cite{chavlemou}, the kinetic theory of point
vortices is more closely related to the kinetic theory of
one-dimensional systems with long-range interactions (like the HMF
model) for which the collision term vanishes identically at order
$1/N$ \cite{bd,cvb,hb2}.

Let us now consider the more general kinetic equation (\ref{gen6}). We
can prove by a direct calculation that this equation conserves the
angular momentum and the linear impulse (see Appendix D of
\cite{kin}). The conservation of the energy is more difficult to
establish by a direct calculation but since Eq.  (\ref{gen6}) is exact
at order $O(1/N)$, the energy must be conserved (the integral
constraints must be conserved at any order). Finally, we note that we
cannot prove the $H$-theorem.  It is only when additional
approximations are made (markovian approximation) that the $H$-theorem
is obtained (see Sec. \ref{sec_ma} and
\cite{chavlemou}).  To be more precise, let us compute the rate of
change of the Boltzmann entropy for point vortices $S=-\int
\frac{\omega}{\gamma}\ln \frac{\omega}{\gamma} d{\bf r}$ with respect 
to the general kinetic equation (\ref{gen6}). After straightforward
manipulations, it can be put in the form
\begin{eqnarray}
\dot S=\frac{1}{2\gamma^{2}}\int d{\bf r}d{\bf r}_{1}\frac{1}{\omega\omega_{1}}\int_{0}^{t}d\tau \biggl\lbrack {\cal V}^{\mu}(1\rightarrow 0)\omega_{1}\frac{\partial\omega}{\partial r^{\mu}}+{\cal V}^{\mu}(0\rightarrow 1)\omega\frac{\partial\omega_{1}}{\partial r_{1}^{\mu}}\biggr\rbrack_{t} \nonumber\\
\times G(t,t-\tau) \biggl\lbrack {\cal V}^{\nu}(1\rightarrow 0)\omega_{1}\frac{\partial\omega}{\partial r^{\nu}}+{\cal V}^{\nu}(0\rightarrow 1)\omega\frac{\partial\omega_{1}}{\partial r_{1}^{\nu}}\biggr\rbrack_{t-\tau}.\label{disc2}
\end{eqnarray}
We note that, because of memory terms, the monotonic increase of the
entropy is not granted. In the case of point vortices, the
decorrelation time is much shorter than the relaxation time (of order
$Nt_{D}$ or larger) so that the markovian approximation is justified
for $N\rightarrow +\infty$. In that case the entropy monotonically
increases as shown explicitly for axisymmetric flows (see
Sec. \ref{sec_ma} and
\cite{chavlemou}). However, as we have already indicated for the
axisymmetric markovian equation (\ref{ma8}), even if $\dot S\ge 0$ and
$\dot E=\dot\Gamma=0$, this does not imply that the kinetic equation
will relax towards the Boltzmann distribution of statistical
equilibrium \cite{chavlemou}.
Indeed, the relaxation can stop before in the absence of resonances. The same remark applies to the more
general equation (\ref{gen6}), valid for non-axisymmetric flows,
although this is more difficult to see. In order to make it clearer,
one possibility would be to use the timescale separation between the
dynamical time $t_{D}$ and the collisional time $t_{coll}\sim Nt_{D}$
and derive an ``orbit-averaged'' kinetic equation in terms of
appropriate variables similar to the angle-action variables used in
other contexts.  In that case, we would get a simpler kinetic
equation, similar to the one derived in \cite{angleaction}, and
exhibiting an appropriate form of ``resonances'' between different
orbits. This would generalize the condition of resonance
$\Omega(r)=\Omega(r')$ associated to Eq. (\ref{ma8}) to the case of
non-axisymmetric flows. Note that a phenomenological
equation, valid for general flows, and exhibiting a form of
``resonances'' required to ensure the conservation of the energy is
provided by Eq. (137) of
\cite{kin} (see also Appendix \ref{sec_heur}).

\subsection{The growth of correlations}
\label{sec_g}

In Sec. \ref{sec_ma}, we have derived the equation satisfied by the
two-body correlation function $g(r_1,r_2,\phi,t)$ at order $1/N$ for
an axisymmetric evolution. To derive the kinetic equation (\ref{ma8}),
we only had to determine the imaginary part of the Fourier transform
of $g$ for $t\rightarrow +\infty$. In this section, we discuss the
growth of the two-body correlation function in more detail. It is
determined by the equation (\ref{ma3}), i.e.:
\begin{eqnarray}
\frac{\partial g}{\partial t}+\Delta\Omega \frac{\partial g}
{\partial \phi}=-\frac{1}{\gamma}\frac{\partial u}{\partial
\phi}\partial {\omega}_1\omega_2.\label{g1}
\end{eqnarray}
We shall assume that the initial vorticity profile is the Boltzmann
distribution of statistical equilibrium 
\begin{eqnarray}
\omega=A e^{-\beta \gamma (\psi+\frac{1}{2}\Omega_L r^2)}.\label{g2}
\end{eqnarray}
This distribution is a stationary solution of Eq. (\ref{ma8}). Therefore,
the r.h.s. of Eq. (\ref{g1}) is independent on time. Using
Eq. (\ref{g2}) and the relations
\begin{eqnarray}
\langle V\rangle_{\theta}(r,t)=-\frac{\partial\psi}{\partial
r}(r,t)=\Omega(r,t)r,\label{g3}
\end{eqnarray}
we find that
\begin{eqnarray}
\partial \omega_1 \omega_2=\beta \gamma \omega_1 \omega_2 \Delta\Omega.\label{g4}
\end{eqnarray}
Substituting this result in Eq. (\ref{g1}) and introducing the function
$h=h({\phi},r_1,r_2,t)$ through the relation $g=\omega_1 \omega_2
h$, we get
\begin{eqnarray}
\frac{\partial h}{\partial t}+\Delta\Omega\frac{\partial h}
{\partial \phi}=-\beta \Delta\Omega \frac{\partial u}{\partial
\phi}.\label{g5}
\end{eqnarray}
Taking the Fourier transform of the foregoing equation and integrating
on time, we obtain
\begin{eqnarray}
\hat{h}_m(t)=-i\beta \int_0^t d\tau m\Delta\Omega \hat{u}_m e^{-i m
\Delta\Omega\tau}=\beta \hat{u}_m (e^{-i m \Delta\Omega
t}-1).\label{g6}
\end{eqnarray}
Therefore
\begin{eqnarray}
\hat{g}_m(r_1,r_2,t)=\beta \hat{u}_m (e^{-i m \Delta\Omega
t}-1)\omega_1\omega_2.\label{g7}
\end{eqnarray}
We note that ${\rm Im}(\hat{g}_{m})$ has a limit (\ref{ma7}) for $t\rightarrow
+\infty$, while ${\rm Re}(\hat{g}_{m})$ has no limit. Taking the inverse
Fourier transform of Eq. (\ref{g7}), we obtain
\begin{eqnarray}
{g}({\phi},{r}_1,{r}_2,t)=\beta \left\lbrack
u(r_1,r_2,{\phi}-\Delta\Omega t)-u(r_1,r_2,{\phi})\right\rbrack
\omega_1 \omega_2.\label{g8}
\end{eqnarray}
This equation describes the growth of two-body correlations in an
axisymmetric flow assuming that the one-body distribution is given by
the Boltzmann distribution. For the potential of interaction
(\ref{pot10}) written in the form
\begin{eqnarray}
u_{12}=-\frac{1}{4\pi}\ln(r_1^2+r_2^2-2r_1 r_2 \cos\phi), \label{g9}
\end{eqnarray}
the correlation function (\ref{g8}) is explicitly given by
\begin{eqnarray}
g(\phi,r_1,r_2,t)=-\frac{\beta}{4\pi}
\ln\left\lbrack\frac{r_1^2+r_2^2-2r_1 r_2 \cos(\phi-\Delta\Omega
t)}{r_1^2+r_2^2-2r_1 r_2 \cos\phi}\right\rbrack \omega_1\omega_2.
\label{g10}
\end{eqnarray}
We note that $g$ has an oscillatory behavior, so that it has no limit
for $t\rightarrow +\infty$. We recall that our approach is valid at
order $1/N$. Therefore, it corresponds to a linear regime extending on
a timescale of order $Nt_{D}$. The two-body correlation function may
reach the value it has at statistical equilibrium on a longer
timescale but next order terms in $1/N$ in the developement must be
taken into account.

\section{Kinetic equations from a quasilinear theory}\label{sec_q}

In this section, we show that the general kinetic equation (\ref{gen6})
describing the {\it collisional} evolution of the point vortex gas can
also be derived from a quasilinear theory of the Klimontovich
equation. We will compare the results with the quasilinear theory of
the 2D Euler equation developed in \cite{prl} to describe the process
of violent relaxation in the {\it collisionless} regime.

\subsection{The slow collisional evolution of point vortices}\label{sec_c}

The exact vorticity profile of a gas of point vortices is a sum of
Dirac functions given by Eq. (\ref{b1}). It satisfies the equation 
\begin{eqnarray}
\frac{\partial \omega_{d}}{\partial t}+{\bf u}_d\nabla\omega_{d}=0,
\label{c1}
\end{eqnarray}
where ${\bf u}_d=-{\bf z}\times\nabla\psi_d$ is the exact velocity
field created by $\omega_d$. Equation (\ref{c1}) is the counterpart of
the Klimontovich equation in plasma physics. It should not be confused
with the 2D Euler equation (\ref{h4ff}) or (\ref{v0}), the counterpart of the
Vlasov equation, which has the same mathematical structure but which
applies to the {\it smooth} vorticity field $\omega$. The 2D Euler
equation is valid during the collisionless regime (see next section)
while the Klimontovich equation is exact and strictly contains the
same information as the Hamiltonian equations (\ref{b2}). We now
decompose the exact vorticity field in the form
$\omega_{d}=\omega+\delta
\omega$ where $\omega=\langle\omega_d\rangle$ is the smooth
vorticity and $\delta\omega$ the fluctuation around it. Substituting
this decomposition in Eq. (\ref{c1}) and locally averaging over the
fluctuations, we get
\begin{eqnarray}
\frac{\partial \omega}{\partial t}+L\omega=-\left\langle \delta{\bf
u}\nabla\delta\omega\right\rangle, \label{c2}
\end{eqnarray}
where $L={\bf u}\cdot \nabla$ is an advection operator constructed with the
smooth velocity field (note that ${\bf u}$ coincides with the field
$\langle {\bf V}\rangle$ introduced previously). Subtracting
Eq. (\ref{c2}) from Eq. (\ref{c1}) and neglecting non linear terms in
the fluctuations, we obtain the following linearized equation for the
evolution of the fluctuations \footnote{As shown in
Sec. \ref{sec_tl}, the proper thermodynamic limit corresponds to
$N\rightarrow +\infty$ in such a way that the individual circulation
$\gamma\sim 1/N$ and the domain area $V\sim 1$. This implies
that $|{\bf r}|\sim 1$. We also have $\omega\sim 1$ and $\delta
\omega\sim 1/\sqrt{N}$ so that $|{\bf u}|\sim 1$ and $\delta |{\bf
u}|\sim 1/\sqrt{N}$. With these scalings, we see that the terms that
we have kept in Eq. (\ref{c3}) are of order $u\delta \omega\sim
{1}/{\sqrt{N}}$ and $\omega\delta u\sim {1}/{\sqrt{N}}$ while
the nonlinear terms that we have neglected are of order $\delta
\omega\delta u\sim {1}/{N}\ll
{1}/{\sqrt{N}}$. We also note that the l.h.s. of Eq. (\ref{c2}) is of
order $\omega\sim 1$ while the r.h.s. of Eq. (\ref{c2}) is of order
$\delta\omega \delta u \sim {1}/{N}$.  Then Eq. (\ref{c2}) can be
rewritten $\partial_{t}{\omega}+L{\omega}=(1/N)C(\omega)$ where the
advective term is of order $O(1)$ and the collision term is of order
$1/N$. Therefore, this equation describes the evolution of the system
on a timescale $\sim Nt_{D}$. For $N\rightarrow +\infty$, it reduces
to the 2D Euler equation $\partial_{t}{\omega}+L{\omega}=0$. In conclusion, the quasilinear theory developed in this section is equivalent to the truncation of the BBGKY hierarchy at the order $1/N$. It amounts to neglecting three-body and higher correlations.}
\begin{eqnarray}
\frac{\partial \delta \omega}{\partial t}+L\delta \omega= -\delta{\bf u}\nabla\omega.
\label{c3}
\end{eqnarray}
Equations (\ref{c2}) and (\ref{c3}) form the basis of the quasilinear theory of
the point vortex gas.  These equations have been studied by Dubin
\& O'Neil \cite{dn}  in the case where the smooth vorticity profile is
axisymmetric.  In that case, Eq. (\ref{c3}) can be solved exactly with
the aid of Laplace-Fourier transforms and a kinetic equation which
takes into account collective effects can be obtained.  This is the
counterpart of the quasilinear theory in plasma physics that is used
to derive the Lenard-Balescu equation from the Klimontovich equation
\cite{pitaevskii}. In the present work, we shall proceed differently
so as to treat the case of systems that are not necessarily
axisymmetric and not necessarily Markovian. Our method avoids the use
of Laplace-Fourier transforms and remains in physical space. This
yields expressions with a clear interpretation which enlightens the
basic physics. The drawback of our approach, however, is that it
neglects collective effects. The formal solution of Eq. (\ref{c3}) is
\begin{eqnarray} \delta \omega(t)=G(t,0)\delta
\omega(0)-\int_{0}^{t}d\tau G(t,t-\tau)\delta{\bf u}
(t-\tau)\nabla\omega(t-\tau), \label{c4}
\end{eqnarray}
where $G$ is the Green function associated with the advection operator
$L$ corresponding to the smooth mean field and we have noted
$\omega(t)=\omega({\bf r},t)$ and $\delta{\bf u}(t)=\delta{\bf u}({\bf
r},t)$ for brevity. On the other hand, the perturbation of the
velocity field is related to the perturbation of the vorticity through
\begin{eqnarray}
\delta {\bf u}(t)=\frac{1}{\gamma}\int {\bf V}(1\rightarrow 0)\delta
\omega_{1}(t)d{\bf r}_{1}, \label{c5}
\end{eqnarray}
where $0$ refers to the position ${\bf r}$. Therefore, considering
Eqs. (\ref{c4}) and (\ref{c5}), we see that the velocity fluctuation
$\delta {\bf u}(t)$ is given by an iterative process: $\delta {\bf
u}(t)$ depends on $\delta \omega_{1}(t)$ which itself depends on
$\delta {\bf u}_{1}(t-\tau)$ etc. We shall solve this problem
perturbatively \footnote{It is at that stage of the
developement that we neglect some collective effects. In the approach
of Dubin \& O'Neil \cite{dn} for axisymmetric flows, Eqs. (\ref{c4})
and (\ref{c5}) can be solved exactly.} in the thermodynamic limit
$N\rightarrow +\infty$ defined in Sec. \ref{sec_tl}. To order $1/N$ we
get
\begin{eqnarray}
\left\langle\delta{\bf u}\nabla\delta\omega \right\rangle=\frac{1}{\gamma}\frac{\partial}{\partial r^{\mu}}\int d{\bf r}_{1}V^{\mu}(1\rightarrow 0)G_{1}(t,0)G(t,0)\langle \delta \omega_{1}(0)\delta \omega(0)\rangle\nonumber\\
-\frac{1}{\gamma^{2}}\frac{\partial}{\partial r^{\mu}}\int_{0}^{t}d\tau\int d{\bf r}_{1}d{\bf r}_{2}V^{\mu}(1\rightarrow 0)
G_{1}(t,t-\tau)G(t,t-\tau)\nonumber\\
\times\biggl\lbrace V^{\nu}(2\rightarrow 0)\langle \delta
\omega_{1}(t-\tau)\delta \omega_{2}(t-\tau)\rangle
\frac{\partial \omega}{\partial r^{\nu}}(t-\tau)\nonumber\\
+ V^{\nu}(2\rightarrow 1)\langle \delta \omega(t-\tau)\delta
\omega_{2}(t-\tau)\rangle \frac{\partial \omega_{1}}{\partial
r_{1}^{\nu}}(t-\tau)\biggr\rbrace. \label{c6}
\end{eqnarray}
 Now, the fluctuation is exactly defined by
\begin{eqnarray}
\delta \omega({\bf r},t)=\sum_{i}\gamma\delta({\bf r}-{\bf
r}_{i}(t))-\omega({\bf r},t). \label{c7}
\end{eqnarray}
Therefore, we obtain
\begin{eqnarray}
\langle \delta \omega_{1}\delta \omega_{2}\rangle =\langle \sum_{i\neq j}\gamma^{2}\delta({\bf r}_{1}-{\bf r}_{i})\delta({\bf r}_{2}-{\bf r}_{j})\rangle +\langle \sum_{i}\gamma^{2}\delta({\bf r}_{1}-{\bf r}_{i})\delta({\bf r}_{2}-{\bf r}_{i})\rangle \nonumber\\
-\langle \sum_{i}\gamma\delta({\bf r}_{1}-{\bf r}_{i})\omega_{2}\rangle-\langle  \sum_{j}\gamma\delta({\bf r}_{2}-{\bf r}_{j})\omega_{1}\rangle+\omega_{1}\omega_{2}.
\label{c8}
\end{eqnarray}
To evaluate the correlation functions, we average with respect to the smooth
distribution $\omega_{i}/(N\gamma)$ or
$\omega_{i}\omega_{j}/(N\gamma)^{2}$. This operation leads to
\begin{eqnarray}
\langle \delta \omega_{1}\delta \omega_{2}\rangle =\frac{N-1}{N}\omega_{1}\omega_{2} +\gamma \omega_{1}\delta({\bf r}_{1}-{\bf r}_{2})
-\omega_{1}\omega_{2}-\omega_{2}\omega_{1}+\omega_{1}\omega_{2},
\label{c9}
\end{eqnarray}
so that, finally,
\begin{eqnarray}
\langle \delta \omega_{1}\delta \omega_{2}\rangle =\gamma \omega_{1}\delta({\bf r}_{1}-{\bf r}_{2})
-\frac{1}{N}\omega_{1}\omega_{2}.
\label{c10}
\end{eqnarray}
Substituting this result in Eq. (\ref{c6}), we obtain
\begin{eqnarray}
\left\langle\delta{\bf u}\nabla\delta\omega \right\rangle=\langle
V^{\mu}(1\rightarrow 0)\rangle\frac{\partial \omega}{\partial
r^{\mu}} +\frac{1}{\gamma}\frac{\partial}{\partial
r^{\mu}}\int_{0}^{t}d\tau\int d{\bf r}_{1}{V}^{\mu}(1\rightarrow 0)
G(t,t-\tau)\nonumber\\
\biggl\lbrace {\cal V}^{\nu}(1\rightarrow
0)\omega_{1}(t-\tau)\frac{\partial \omega}{\partial r^{\nu}}(t-\tau)
+ {\cal V}^{\nu}(0\rightarrow 1)\omega(t-\tau)\frac{\partial
\omega_{1}}{\partial r_{1}^{\nu}}(t-\tau)\biggr\rbrace, \label{c11}
\end{eqnarray}
where we have regrouped the two Greenians $G$ and $G_1$ in a single
notation for brevity. Finally, replacing this expression in Eq. (\ref{c2}),
we obtain the kinetic equation
\begin{eqnarray}
\frac{\partial \omega}{\partial t}+\frac{N-1}{N} {\bf
u}\nabla\omega=\gamma\frac{\partial}{\partial
r^{\mu}}\int_{0}^{t}d\tau\int d{\bf r}_{1}\frac{{\cal
V}^{\mu}}{\gamma}(1\rightarrow 0)
G(t,t-\tau)\nonumber\\
\times \biggl\lbrace \frac{{\cal V}^{\nu}}{\gamma}(1\rightarrow
0)\omega_{1}\frac{\partial\omega}{\partial r^{\nu}} + \frac{{\cal
V}^{\nu}}{\gamma}(0\rightarrow 1)\omega \frac{\partial\omega_{1}}{\partial
r_{1}^{\nu}}\biggr\rbrace_{t-\tau}. \label{c12}
\end{eqnarray}
This is identical to the general kinetic equation (\ref{gen6}) obtained
from the projection operator formalism or from the  BBGKY-like
hierarchy. We note that the term of order $1/N$ in the l.h.s. comes
from the first term  in Eq. (\ref{c4}). It corresponds to the mere advection
of the vorticity fluctuation by the smooth velocity field in Eq. (\ref{c3}), i.e. ignoring the coupling between the velocity fluctuations
and the smooth vorticity (r.h.s. of Eq. (\ref{c3})) which gives rise to the
collision term.

\subsection{The violent collisionless evolution of point vortices}
\label{sec_v}

To leading order in $N\rightarrow +\infty$, the smooth vorticity
profile of the point vortex gas is solution of the 2D Euler-Poisson
system 
\begin{eqnarray}
\frac{\partial \omega}{\partial t}+{\bf
u}\nabla\omega=0, \qquad \omega=-\Delta\psi. \label{v0}
\end{eqnarray} 
This is the counterpart of the Vlasov-Poisson system in stellar
dynamics and plasma physics. The 2D Euler equation describes the
collisionless evolution of the point vortices due to mean field
effects before collisions come into play on a timescale $Nt_D$ or
larger. Starting from an initial condition which is dynamically
unstable, the 2D Euler-Poisson system develops an intricate
filamentation at smaller and smaller scales. In this sense, the
fine-grained vorticity $\omega({\bf r},t)$ never achieves
equilibrium. However, if we locally average over the filaments, the
resulting ``coarse-grained'' vorticity $\overline{\omega}({\bf r},t)$
will achieve a steady state on a timescale $\sim t_{D}$
\cite{houches}. Since the 2D Euler equation is only valid in the
collisionless regime $t\ll t_{coll}$, this corresponds to a
quasi-stationary state (QSS) that will slowly evolve due to the effect
of collisions occuring on a longer timescale $\sim Nt_{D}$ or
larger. We can try to predict this QSS in terms of a statistical
mechanics of the 2D Euler equation, using the approach of
Miller-Robert-Sommeria \cite{miller,rs}
\footnote{In these works, the 2D Euler equation 
is justified as a limit of the Navier-Stokes equation for inviscid
fluids $\nu\rightarrow 0$. Since the 2D Euler equation also describes
the collisionless regime of the point vortex gas, their approach can
be applied in that context.}. This is the 2D hydrodynamic version of
the theory of violent relaxation proposed by Lynden-Bell \cite{lb} for
collisionless stellar systems based on the Vlasov equation. In the
case where the fine-grained vorticity $\omega({\bf r},t)$ takes only
two values $0$ and $\sigma_{0}$, the statistical equilibrium state
maximizes the mixing entropy
\begin{eqnarray}
S_{MRS}=-\int \biggl\lbrace \frac{\overline{\omega}}{\sigma_{0}}\ln \frac{\overline{\omega}}{\sigma_{0}}+\biggl (1- \frac{\overline{\omega}}{\sigma_{0}}\biggr )\ln \biggl (1-\frac{\overline{\omega}}{\sigma_{0}}\biggr )\biggr\rbrace d{\bf r},
\label{v1}
\end{eqnarray}
at fixed circulation and energy. This leads to the coarse-grained
vorticity 
\begin{eqnarray}
\overline{\omega}=\frac{\sigma_{0}}{1+\lambda e^{\beta\sigma_{0}\psi}}.
\label{v2}
\end{eqnarray}
Note that the mixing entropy (\ref{v1}) is formally similar to the
Fermi-Dirac entropy and the equilibrium distribution (\ref{v2}) is
formally similar to the Fermi-Dirac distribution. An effective
``exclusion principle'', similar to the Pauli principle in quantum
mechanics, arises in the theory of violent relaxation of continuous
vorticity fields because the different levels of vorticity cannot
overlap. We also note that the violent relaxation of point vortices
does not lead to a segregation of the vortices according to their
circulation (in the multi-species case) because the circulation of the
individual vortices does not appear in the 2D Euler equation. Finally,
we stress that the MRS theory is based on an assumption of
ergodicity. Indeed, it implicitly assumes that the vorticity mixes
well so that the QSS is the {\it most mixed state} compatible with the
integral constraints of the 2D Euler equation. This may not always be
the case as discussed in Sec. \ref{sec_ivr}.

A kinetic theory of the process of violent relaxation has been
developed in \cite{prl} with the aim to determine the dynamical
equation satisfied by the coarse-grained vorticity field
$\overline{\omega}({\bf r},t)$. This approach is based on a
quasilinear approximation of the 2D Euler equation that is formally
similar to that developed in the previous section (but with a
completely different interpretation). In Sec. \ref{sec_c}, the
subdynamics was played by $\omega_{d}$ (a sum of $\delta$-functions)
and the macrodynamics by $\omega$ (a smooth field). The smooth field
averages over the positions of the $\delta$-functions (point
vortices) that strongly fluctuate. In the phase of violent relaxation,
the ``smooth'' field $\omega$ develops itself a finely striated
structure and strongly fluctuates. Therefore, it is {\it not} smooth
at a higher scale of resolution and a second smoothing procedure
(coarse-graining) must be introduced. In that case, the subdynamics is
played by $\omega$ and the macrodynamics by $\overline{\omega}$. The
coarse-grained field averages over the positions of the filaments. The
quasilinear theory leads to a kinetic equation for the coarse-grained
vorticity of the form \cite{prl}:
\begin{eqnarray}
\frac{\partial \overline{\omega}}{\partial t}+\overline{\bf u}\nabla\overline{\omega}=\epsilon^{2}\frac{\partial}{\partial r^{\mu}}\int_{0}^{t}d\tau\int d{\bf r}_{1}\frac{{V}^{\mu}}{\gamma}(1\rightarrow 0)
G(t,t-\tau)\nonumber\\
\times \biggl\lbrace \frac{{V}^{\nu}}{\gamma}(1\rightarrow 0)\overline{\omega}_{1}(\sigma_{0}-\overline{\omega}_{1})\frac{\partial\overline{\omega}}{\partial r^{\nu}}
+ \frac{{V}^{\nu}}{\gamma}(0\rightarrow 1)\overline{\omega}(\sigma_{0}-\overline{\omega})\frac{\partial\overline{\omega}_{1}}{\partial r_{1}^{\nu}}\biggr\rbrace_{t-\tau},
\label{v3}
\end{eqnarray}
where $\epsilon$ is the coarse-graining mesh size and $\overline{\bf
u}$ the velocity field produced by the coarse-grained vorticity
(recall also that the ratio ${\bf V}(1\rightarrow 0)/\gamma$ only
depends on ${\bf r}_{1}$ and ${\bf r}$ according to Eq. (\ref{b5})).
This equation is expected to describe the late quiescent stages of the
relaxation process when the fluctuations have weaken so that the
quasilinear approximation can be implemented. It does not describe the
early, very chaotic, process of violent relaxation driven by the
strong fluctuations of the stream function. The quasilinear theory of
the 2D Euler equation is therefore a theory of ``quiescent''
collisionless relaxation. This is the counterpart of the quasilinear
theory of the Vlasov-Poisson system developed for collisionless
stellar systems
\cite{kp,sl,dub}.

Equation (\ref{v3}) is very similar, in structure, to Eq. (\ref{c12})
for the collisional evolution of point vortices, with nevertheless
three important differences: (i) the fluctuating velocity ${\cal
V}(1\rightarrow 0)$ is replaced by the direct velocity
${V}(1\rightarrow 0)$ because the fluctuations are taken into account
differently. (ii) The vorticity $\omega$ in the collisional term of
equation (\ref{c12}) is replaced by the product
$\overline{\omega}(\sigma_{0}-\overline{\omega})$ in equation
(\ref{v3}). This nonlinear term arises from the effective ``exclusion
principle'' accounting for the non-overlapping of vortex patches in
the collisionless regime. This is consistent with the Fermi-Dirac-like
entropy (\ref{v1}) and Fermi-Dirac-like distribution (\ref{v2}) at
statistical equilibrium. (iii) Considering the dilute limit
$\overline{\omega}\ll\sigma_{0}$ to fix the ideas, we see that the
equations (\ref{v3}) and (\ref{c12}) have the same mathematical form
differing only in the prefactors: the circulation $\gamma$ of a point
vortex in Eq.  (\ref{c12}) is replaced by the circulation
$\sigma_{0}\epsilon^{2}$ of a completely filled macrocell in
Eq. (\ref{v3}). This implies that the timescales of collisional and
collisionless ``relaxation'' are in the ratio
\begin{eqnarray}
\frac{t_{ncoll}}{t_{coll}}\sim \frac{\gamma}{\sigma_{0}\epsilon^{2}}.
\label{v4}
\end{eqnarray}
Since $\sigma_{0}\epsilon^{2}\gg \gamma$, this ratio is in general
quite small implying that the collisionless relaxation is much more
rapid than the collisional relaxation. Typically, $t_{ncoll}$ is of
the order of a few dynamical times $t_{D}$ (its
precise value depends on the size of the mesh) while $t_{coll}$ is of
order $Nt_{D}$, or larger. The kinetic equation (\ref{v3}) 
conserves the circulation, the angular momentum and, presumably, the energy. By
contrast, we cannot prove an $H$-theorem for the MRS entropy
(\ref{v1}). Indeed, the time variation of the MRS entropy is of the
form
\begin{eqnarray}
\dot S_{MRS}=\frac{1}{2}\epsilon^{2}\int d{\bf r}d{\bf r}_{1}\frac{1}{\overline{\omega}(\sigma_{0}-\overline{\omega})\overline{\omega}_{1}(\sigma_{0}-\overline{\omega}_{1})}\int_{0}^{t}d\tau Q(t) G(t,t-\tau)Q(t-\tau),\label{g6fgh}
\end{eqnarray}
\begin{eqnarray}
Q(t)= \left\lbrack \frac{{V}^{\mu}}{\gamma}(1\rightarrow 0)\overline{\omega}_{1}(\sigma_{0}-\overline{\omega}_{1})\frac{\partial \overline{\omega}}{\partial r^{\mu}}+ \frac{{V}^{\mu}}{\gamma}(0\rightarrow 1)\overline{\omega}(\sigma_{0}-\overline{\omega}) \frac{\partial \overline{\omega}_{1}}{\partial r_{1}^{\mu}}\right\rbrack,\label{g6fghb}
\end{eqnarray}
and its sign is not necessarily positive. This depends on the
importance of memory effects.

\subsection{Discussion: incomplete violent relaxation}\label{sec_ivr}

Even if Eq. (\ref{v3}) conserves the energy and the circulation and
increases the MRS entropy (\ref{v1}) monotonically, this does
not necessarily imply that the system will converge towards the MRS
distribution (\ref{v2}). There are many cases where the MRS theory
provides a good prediction of the QSS \cite{rsmepp,staquet}. However,
it has also been observed in some experiments \cite{hd} and numerical
simulations \cite{brands} that the QSS does not exactly coincide with
the strict statistical equilibrium state predicted by the MRS theory
because of the complicated problem of {\it incomplete relaxation}
\cite{next05}. This is usually explained by a lack of ergodicity or
``incomplete mixing''. Here, we try to be a little more precise by
using the kinetic theory. There can be several reasons of incomplete
relaxation:

(i) {\it Absence of resonances:} Very few is known concerning
kinetic equations of the form of Eq. (\ref{v3}) and it is not clear whether
the MRS distribution (\ref{v2}) is a stationary solution of that
equation (and whether it is the only one). In order to simplify
Eq. (\ref{v3}), we shall assume that the coarse-grained vorticity
field is axisymmetric and that the correlations relax on a timescale
which is much shorter than the typical time on which the
coarse-grained vorticity field changes (Markovian
approximation). Although this approximation was justified to describe
the collisional relaxation of point vortices (because of the timescale
separation between the dynamical time and the collision time), this
approximation is not clear for the process of violent collisionless
relaxation where memory terms can be important. However, with this
approximation we can re-do the calculations of the previous sections
and obtain a kinetic equation of the form
\begin{eqnarray}
\frac{\partial \overline{\omega}}{\partial
t}=-\frac{\epsilon^{2}}{4r}  \frac{\partial}{\partial r}
\int_0^{+\infty} r_1 dr_1 \ln\left\lbrack 1-\left
(\frac{r_<}{r_>}\right )^{2}\right\rbrack\delta(\Omega-\Omega_1)
\left \lbrace\frac{1}{r}\omega_1(\sigma_{0}-\omega_1)\frac{\partial
\omega}{\partial r}-\frac{1}{r_1}\omega
(\sigma_{0}-\omega)\frac{\partial \omega_1}{\partial
 r_1}\right \rbrace.\nonumber\\ \label{v5}
\end{eqnarray}
This equation conserves the circulation, the energy and the angular
momentum and satisfies an H-theorem for the mixing entropy (\ref{v1}).
However, as already discussed for the collisional evolution of the
point vortices, it does {\it not} relax in general towards the
statistical equilibrium state, here the MRS distribution
(\ref{v2}) because of the absence of resonances. The same conclusion
probably applies to the more general equation (\ref{v3}) and to the
heuristic equation (148) of \cite{kin} which also conserve $E$,
$\Gamma$, $L$ and increase the mixing entropy (\ref{v1}) but do not
obligatory relax towards the MRS distribution (\ref{v2}). The
system tries to approach the statistical equilibrium state (as
indicated by the increase of the entropy) but may be trapped in a QSS
that is different from the statistical prediction (\ref{v2}). This QSS is a
steady solution of Eq.  (\ref{v3}) which cancels individually the advective
term (l.h.s.) and the effective collision term (r.h.s.).  This
determines a subclass of steady states of the 2D Euler equation
(cancellation of the l.h.s.)  such that the complicated ``turbulent''
current ${\bf J}$ in the r.h.s. vanishes.  This offers a large class
of possible steady state solutions that can explain the deviation
between the QSS and the MRS statistical equilibrium state
(\ref{v2}) observed, in certain cases, in simulations and experiments
of violent relaxation.  
One may argue that nonlinear terms are
needed in the kinetic theory in order to obtain an equation that
relaxes towards the statistical equilibrium distribution
(\ref{v2}). In that case, we must develop a kinetic theory that goes
beyond the quasilinear approximation.

(ii) {\it Incomplete
relaxation in space:} The turbulent current ${\bf J}$ in Eq. (\ref{v3}) is
driven by the fluctuations $\omega_{2}\equiv
\overline{\tilde{\omega}^{2}}=\overline{\omega^{2}}-\overline{\omega}^{2}$ of the vorticity (generating the
fluctuations $\delta {\bf u}$ of the velocity) \cite{prl}. In the
``mixing region'' where the fluctuations are strong, the vorticity
tends to reach the MRS distribution (\ref{v2}). As we depart from the
``mixing region'', the fluctuations decay ($\omega_{2}\rightarrow 0$)
and the mixing is less and less efficient. In these regions, the
system takes a long time to reach the MRS distribution (\ref{v2}) and,
in practice, cannot attain it in the time available (see (iii)). In
the two levels case, we have
$\omega_{2}=\overline{\omega}(\sigma_{0}-\overline{\omega})$. Therefore,
the regions where $\overline{\omega}\rightarrow 0$ or
$\overline{\omega}\rightarrow \sigma_{0}$ do not mix well (the
diffusion current ${\bf J}$ is weak) and the observed vorticity can be
sensibly different from the MRS distribution in these regions.  This
concerns essentially the core ($\overline{\omega}\rightarrow
\sigma_{0}$) and the tail ($\overline{\omega}\rightarrow 0$) of the
vorticity distribution. This result, derived from the kinetic theory,
is consistent with what is observed in experiments
\cite{hd,brands}: the core vorticity decreases less than predicted by
the MRS statistical theory while the tail of the vorticity profile
decreases more rapidly. For the same reason, {\it it can also explain why
the vorticity peak is remarkably well conserved during a merging
process} as observed in 2D decaying turbulence \cite{carnevale}.

(iii) {\it Incomplete relaxation in time:} during violent relaxation,
the system tends to approach the statistical equilibrium state
(\ref{v2}). However, as it approaches equilibrium, the fluctuations of
the velocity field, which are the engine of the evolution, become less
and less effective to drive the relaxation. This is because the scale
of the fluctuations becomes smaller and smaller as time goes on. This
effect can be taken into account in the kinetic theory by considering
that the correlation length $\epsilon(t)$ decreases with time so that,
in the kinetic equation (\ref{v3}), the prefactor
$\epsilon(t)\rightarrow 0$ rapidly.  As a result, the ``turbulent''
current ${\bf J}$ in Eq. (\ref{v3}) can vanish {\it before} the system
has reached the statistical equilibrium state (\ref{v2}). In that
case, the system can be trapped in a QSS that is a steady solution of
the 2D Euler equation different from the statistical prediction
(\ref{v2}).

\section{Relaxation of a test vortex in a bath}\label{sec_r}

In this section, we study the relaxation of a test vortex in a bath of
field vortices. Specifically, we consider a collection of $N$ point
vortices at statistical equilibrium (thermal bath) and introduce a new
vortex in the system. To leading order in $N\rightarrow +\infty$, the
point vortex is advected by the mean flow. However, due to finite $N$
effects, the test vortex undergoes discrete interactions with the
vortices of the bath and progressively acquires their distribution. We
wish to study this stochastic process. The probability density $P({\bf
r},t)$ of finding the test vortex in ${\bf r}$ at time $t$ is governed
by a Fokker-Planck equation involving a term of diffusion and a term
of drift. In our previous papers, we obtained these terms from a
linear response theory \cite{preR} or from the projection operator
formalism \cite{kin}. In the present work, we obtain these terms
directly from the equations of motion and show how collective effects
can be included in the theory in the case of axisymmetric flows. Our
approach is also valid if the bath is made of a distribution of field
vortices that evolves {\it slowly}, so that it can be assumed
stationary on a timescale $(N/\ln N)t_D$ which is the typical
relaxation time of the test vortex in the bath (see \cite{chavlemou}
for details).

\subsection{Diffusion coefficient}\label{sec_d}

The increment of the position of the test vortex between $t$ and
$t-s$ due to the fluctuations of the velocity is
\begin{eqnarray}
\Delta r^{\mu}=\int_{t-s}^{t}{\cal V}^{\mu}(t')dt'.
\label{d1}
\end{eqnarray}
After standard calculations (see, e.g., Sec. 4.2 of \cite{curious}),
the second moment of the increments of position can be rewritten in
the form
\begin{eqnarray}
\left\langle \frac{\Delta r^{\mu}\Delta r^{\nu}}{2s}\right\rangle=\frac{1}{s}\int_{0}^{s}(s+\tau)\langle {\cal V}^{\mu}(t){\cal V}^{\nu}(t-\tau)\rangle d\tau.
\label{d2}
\end{eqnarray}
We shall assume that the correlation function decreases more rapidly
than $\tau^{-1}$. Then, taking the limit $s\rightarrow +\infty$, we
find that the diffusion coefficient is given by the Kubo formula
\begin{eqnarray}
D^{\mu\nu}=\left\langle \frac{\Delta r^{\mu}\Delta r^{\nu}}{2\Delta
t}\right\rangle=\lim_{s\rightarrow +\infty}\left\langle \frac{\Delta
r^{\mu}\Delta r^{\nu}}{2s}\right\rangle=\int_{0}^{+\infty}\langle
{\cal V}^{\mu}(t){\cal V}^{\nu}(t-\tau)\rangle d\tau. \label{d3}
\end{eqnarray}
On the other hand, after straightforward calculations (see, e.g.,
Sec. 4.1 of \cite{curious}), we obtain
\begin{eqnarray}
\langle {\cal V}^{\mu}(t){\cal V}^{\nu}(t-\tau)\rangle=N\langle {\cal V}^{\mu}(1\rightarrow 0,t){\cal V}^{\nu}(1\rightarrow 0,t-\tau)\rangle\nonumber\\
=\int d{\bf r}_{1}{\cal V}^{\mu}(1\rightarrow 0,t){\cal
V}^{\nu}(1\rightarrow 0,t-\tau)\frac{\omega}{\gamma}({\bf r}_{1}).
\label{d4}
\end{eqnarray}
Therefore, combining Eqs. (\ref{d3}) and (\ref{d4}), we get
\begin{eqnarray}
D^{\mu\nu}=\int_{0}^{+\infty} d\tau d{\bf r}_{1}{\cal
V}^{\mu}(1\rightarrow 0,t) {\cal V}^{\nu}(1\rightarrow
0,t-\tau)\frac{\omega}{\gamma}({\bf r}_{1}). \label{d5}
\end{eqnarray}
For an axisymmetric system, the diffusion coefficient is given by
\begin{eqnarray}
D=\left\langle \frac{(\Delta r)^{2}}{2\Delta
t}\right\rangle=\lim_{s\rightarrow +\infty}\left\langle
\frac{(\Delta r)^{2}}{2s}\right\rangle=\int_{0}^{+\infty}\langle
{V}_{r}(t){V}_{r}(t-\tau)\rangle d\tau, \label{d6}
\end{eqnarray}
leading to
\begin{eqnarray}
D=\int_{0}^{+\infty} d\tau
\int_{0}^{2\pi}d\theta_{1}\int_{0}^{+\infty}
r_{1}d{r}_{1}{V}_{r}(1\rightarrow 0,t){V}_{r}(1\rightarrow
0,t-\tau)\frac{\omega}{\gamma}({r}_{1}). \label{d7}
\end{eqnarray}
If we neglect collective effects, the velocity created by a field
vortex $1$ on the test vortex $0$ is given by (see Appendix \ref{sec_pot}):
\begin{eqnarray}
{V}_{r}(1\rightarrow 0,t)=i\gamma \frac{1}{r}\sum_{m} m
\hat{u}_{m}e^{i m ({\theta}-{\theta}_{1})}. \label{d8}
\end{eqnarray}
At time $t-\tau$, we have
\begin{eqnarray} {V}_{r}(1\rightarrow
0,t-\tau)=i\gamma \frac{1}{r}\sum_{m} m \hat{u}_{m}e^{i m
({\theta}(t-\tau)-{\theta}_{1}(t-\tau))}. \label{d9}
\end{eqnarray}
To leading order in $N\rightarrow +\infty$, the point
vortices are advected by the  mean field velocity  so that
${\theta}_i(t-\tau)={\theta}_i-{\Omega(r_i)}\tau$ and
$r_i(t-\tau)=r_i$ where $r_i=r_i(t)$ and $\theta_i=\theta_i(t)$
denote their position at time $t$. Thus, we get
\begin{eqnarray}
{V}_{r}(1\rightarrow 0,t-\tau)=i\gamma \frac{1}{r}\sum_{m} m
\hat{u}_{m}e^{i m (\phi-\Delta\Omega \tau)}. \label{d10}
\end{eqnarray}
Substituting this expression in Eq. (\ref{d7}) and carrying out the
integrations on $\theta_1$ and $\tau$, we obtain after
straightforward calculations
\begin{eqnarray}
D=\frac{2\pi^{2}\gamma}{r^{2}}\int_{0}^{+\infty}r_{1}dr_{1}\chi(r,r_{1})\delta(\Delta\Omega)\omega({r}_{1}),\label{d11}
\end{eqnarray}
where the function $\chi(r,r_{1})$ is defined in Eq. (\ref{ma9}).  If the
profile of angular velocity is monotonic, we can use
$\delta(\Delta\Omega)=\delta(r-r_1)/|\Omega'(r)|$ and we find that
\begin{eqnarray}
D(r)=2\pi^{2}\gamma\frac{\chi(r,r)}{|\Sigma(r)|}\omega(r),\label{d12}
\end{eqnarray}
where $\Sigma=r\Omega'(r)$ is the local shear. For the potential (\ref{pot10}), we have
\begin{eqnarray}
\chi(r,r)=\frac{1}{8\pi^2}\sum_{m=1}^{+\infty}\frac{1}{m}=\frac{1}{8\pi^2}\ln\Lambda,\label{d13}
\end{eqnarray}
where $\ln\Lambda\equiv\sum_{m=1}^{+\infty}\frac{1}{m}$ is a
logarithmically diverging Coulomb factor that has to be
regularized. This leads to the following expression of the diffusion
coefficient
\begin{eqnarray}
D(r)=\frac{\gamma}{4}\ln\Lambda
\frac{1}{|\Sigma(r)|}\omega(r).\label{d14}
\end{eqnarray}
This expression, with the shear reduction, was derived in Chavanis
\cite{preR,kin} and Dubin \& Jin \cite{jin} from the Kubo formula. Because
of the divergence of the Coulomb factor, the value of the diffusion
coefficient is dominated by the contribution of field vortices at
radial distance $r_1=r$, justifying the {\it local approximation} made in
\cite{preR}. Therefore, Eq. (\ref{d14}) gives the {\it dominant term} in the diffusion coefficient (\ref{d11}) of a test vortex even if the profile of
angular velocity of the field vortices is non-monotonic. In practice,
the Coulomb factor has to be regularized as discussed in detail in
\cite{jin,chavlemou}. It is then found that $\ln\Lambda$ scales with the number
of particles like $\frac{1}{2}\ln N$ in the thermodynamic limit
$N\rightarrow +\infty$, in agreement with the rough estimates 
in \cite{preR}.

\subsection{Drift coefficient}\label{sec_dc}

In addition to its diffusive motion, a test vortex immersed in a
bath of field vortices with spatially inhomogeneous vorticity
distribution undergoes a systematic drift. The drift corresponds to
the response of the field vortices to the perturbation caused by the
test vortex, as in a polarization process. The test vortex modifies
the density distribution  of the field vortices and the retroaction
of this perturbation on  the test vortex causes its drift. The
expression of the drift can be derived from a linear response theory
starting from the Liouville equation as done in \cite{preR}. In this section,
we show that it can also be obtained from the Klimontovich equation.
This will make a close connection with the quasilinear theory
developed in Sec. \ref{sec_q}.

The introduction of a test vortex in a bath of field vortices modifies
the vorticity profile $\omega({\bf r},t)$ of the bath. Since this
perturbation is small, it can be described by the linearized Euler
equation
\begin{eqnarray}
\frac{\partial \delta \omega}{\partial t}+L\delta \omega=
-\delta{\bf u}\nabla \omega, \label{dc1}
\end{eqnarray}
whose  formal solution is
\begin{eqnarray}
\delta \omega(t)=-\int_{0}^{t}d\tau
G(t,t-\tau)\delta{\bf u} (t-\tau)\nabla\omega(t-\tau). \label{dc2}
\end{eqnarray}
We have used the fact that, initially, $\delta\omega(0)=0$.
On the other hand, the perturbation of the velocity field in ${\bf r}$
is given by
\begin{eqnarray}
\delta{\bf u}({\bf r},t)=\frac{1}{\gamma}\int {\bf V}(1\rightarrow 0)\delta
\omega_{1}(t)d{\bf r}_{1} +\int {\bf \cal V}(1\rightarrow 0)\delta({\bf r}_{1}-{\bf r}_{P}(t)) d{\bf r}_{1}. \label{dc3}
\end{eqnarray}
The second term is the velocity created by the test vortex at position
${\bf r}_{P}(t)$ and the first term is the fluctuation of the velocity due
to the perturbed density distribution of the field vortices.
Substituting Eq. (\ref{dc2}) in Eq. (\ref{dc3}), we obtain
\begin{eqnarray}
\delta {\bf u}({\bf r},t)=-\frac{1}{\gamma}\int_{0}^{t}d\tau\int d{\bf r}_{1}{\bf V}(1\rightarrow 0)
G_{1}(t,t-\tau)\delta u_{1}^{\nu}(t-\tau)\frac{\partial \omega_{1}}{\partial r_{1}^{\nu}} (t-\tau)\nonumber\\
+\int {\bf \cal V}(1\rightarrow 0)\delta({\bf r}_{1}-{\bf r}_{P}(t)) d{\bf r}_{1}.
\label{dc4}
\end{eqnarray}
This is an integral equation for $\delta {\bf u}({\bf r},t)$. For an
axisymmetric flow, this equation can be solved exactly by using
Laplace-Fourier transforms as done in Schecter \& Dubin
\cite{schecter}. In order to treat more general flows, we shall make
an approximation which amounts to neglecting some collective
effects. We solve Eq. (\ref{dc4}) by an iterative process: we first
neglect the first term in Eq. (\ref{dc4}) keeping only the
contribution of the test particle. Then, we substitute this value in
the first term of the r.h.s of Eq. (\ref{dc4}). This operation gives
\begin{eqnarray}
\delta {\bf u}({\bf r},t)=-\frac{1}{\gamma}\int_{0}^{t}d\tau\int d{\bf r}_{1}d{\bf r}_{2}{\bf V}(1\rightarrow 0)G_{1}(t,t-\tau){\cal V}^{\nu}(2\rightarrow 1)\nonumber\\
\times\frac{\partial \omega_{1}}{\partial r_{1}^{\nu}}
(t-\tau)\delta({\bf r}_{2}-{\bf r}_{P}(t-\tau)) +\int {\bf \cal
V}(1\rightarrow 0)\delta({\bf r}_{1}-{\bf r}_{P}(t))d{\bf r}_{1}.
\label{dc5}
\end{eqnarray}
This quantity represents the fluctuation of the velocity field in
${\bf r}$ caused by the introduction of a test vortex in the system
and taking into account the retroaction of the field vortices. If we
evaluate this expression at the position ${\bf r}_{P}$ of the test
vortex and subtract the second term (self-interaction), we obtain the
drift experienced by the test vortex in response to the perturbation
that it caused. Denoting finally by $0$ the position of the test vortex, we
find that its drift is given by
\begin{eqnarray}
V^{\mu}_{pol}=-\frac{1}{\gamma}\int_{0}^{t}d\tau\int d{\bf
r}_{1}{V}^{\mu}(1\rightarrow 0,t){\cal
V}^{\nu}(0\rightarrow 1,t-\tau)\frac{\partial \omega}{\partial r_1^{\nu}}
({\bf r}_{1}(t-\tau)). \label{dc6}
\end{eqnarray}
For a thermal bath, where the distribution of the field vortices is given by  $\omega({\bf r}_{1})=A e^{-\beta \gamma \psi({\bf r}_{1})}$, we obtain
\begin{eqnarray}
V^{\mu}_{pol}=\beta\int_{0}^{t}d\tau\int d{\bf
r}_{1}{V}^{\mu}(1\rightarrow 0,t){\cal
V}(0\rightarrow 1,t-\tau)\cdot \nabla\psi({\bf r}_{1}(t-\tau))\omega({\bf r}_{1}), \label{dc6bis}
\end{eqnarray}
where we have used $\omega({\bf r}_{1}(t-\tau))=\omega({\bf
r}_{1}(t))$ since $\omega$ is a stationary solution of the 2D Euler
equation. This is equivalent to the result of the linear response
theory based on the Liouville equation \cite{preR} but it is obtained
here in a simpler manner.  We can also obtain this result in a
slightly different way. We approximate $\delta {\bf u}({\bf r},t)$ in
Eq. (\ref{dc1}) by the velocity ${\cal V}(P\rightarrow 0)$ created by
the test vortex so that
\begin{eqnarray}
\frac{\partial \delta \omega}{\partial t}+L\delta \omega=
-{\cal V}(P\rightarrow 0)\nabla \omega. \label{dc7}
\end{eqnarray}
This equation can be solved with Green functions yielding
\begin{eqnarray}
\delta \omega(t)=
-\int_{0}^{t}d\tau G(t,t-\tau) {\cal V}(P\rightarrow 0,t-\tau)\nabla \omega(t-\tau). \label{dc8}
\end{eqnarray}
This represents the perturbation of the distribution of field vortices
caused by the introduction of the test vortex in the system. This
perturbation produces in turn a velocity which causes the drift of the
test vortex (by retroaction). If we substitute Eq. (\ref{dc8}) in the
first part of Eq. (\ref{dc3}) and evaluate this quantity at the
position of the test vortex, we recover Eq. (\ref{dc6}) for the drift
(see also Appendix \ref{sec_da}).

If we now consider an axisymmetric distribution of field vortices, the expression of the drift becomes
\begin{eqnarray}
V_{r}^{pol}=-\frac{1}{\gamma}\int_{0}^{t}d\tau\int_{0}^{2\pi}d\theta_{1}\int_{0}^{+\infty}
r_{1}dr_{1}{V}_{r}(1\rightarrow 0,t){V}_{r_1}(0\rightarrow
1,t-\tau)\frac{d\omega}{d r} ({r}_{1}). \label{dc9}
\end{eqnarray}
Using the identity (\ref{pot5}) and taking the limit $t\rightarrow +\infty$,
we get
\begin{eqnarray}
V_{r}^{pol}=\frac{1}{\gamma}\int_{0}^{+\infty}d\tau
\int_{0}^{2\pi}d\theta_{1}\int_{0}^{+\infty}r_{1}dr_{1}\frac{r}{r_{1}}{V}_{r}(1\rightarrow
0,t){V}_{r}(1\rightarrow 0,t-\tau)\frac{d\omega}{d r} ({r}_{1}).
\label{dc10}
\end{eqnarray}
This is a sort of generalized Kubo relation involving the {\it gradient} of
the density profile instead of the density profile itself. {\it The nice
similarity in the expressions of the diffusion coefficient
(\ref{d7}) and drift term (\ref{dc10}) is worth mentioning}. The
integrals on ${\theta}_{1}$ and $\tau$ can be evaluated in the same
manner as in Sec. \ref{sec_d} and we obtain
\begin{eqnarray}
V_{r}^{pol}=\frac{2\pi^{2}\gamma}{r}\int_{0}^{+\infty}dr_{1}\chi(r,r_{1})
\delta(\Delta\Omega)\frac{d \omega}{d r} ({r}_{1}).\label{dc11}
\end{eqnarray}
Now, the drift of the test vortex is due not only to the
polarization process but also to the variation of the diffusion
coefficient with ${r}$. As a result, the complete expression of the
drift is
\begin{eqnarray}
V_{r}^{drift}\equiv \left\langle \frac{\Delta r}{\Delta
t}\right\rangle=\frac{\partial D}{\partial r}+V_{r}^{pol}.
\label{dc12}
\end{eqnarray}
From  Eqs. (\ref{d11}) and (\ref{dc11}), we obtain
\begin{eqnarray}
V_{r}^{drift}=2\pi^2\gamma\int_0^{+\infty}r r_1 dr_1
\omega(r_1)\left (\frac{1}{r}\frac{\partial}{\partial
r}-\frac{1}{r_1}\frac{\partial}{\partial r_1}\right
)\chi(r,r_1)\delta(\Delta\Omega)\frac{1}{r^2}, \label{dc13}
\end{eqnarray}
where we have used an integration by parts in Eq. (\ref{dc11}). Expressions (\ref{d11}) and (\ref{dc13}) for the diffusion coefficient and the drift term can also be
obtained directly from the Hamiltonian equations, by making a
systematic expansion of the trajectory of the point vortices in
powers of $1/N$ in the limit $N\rightarrow +\infty$ as shown in
Appendix C of \cite{chavlemou}.

For a thermal bath, corresponding to the case where the field
vortices are at statistical equilibrium, the vorticity profile is
given by the Boltzmann distribution
\begin{eqnarray}
\omega({r}_{1})=Ae^{-\beta \gamma \psi'(r_1)}, \label{dc14}
\end{eqnarray}
where $\psi'=\psi+(1/2)\Omega_L r^2$ is the relative stream
function. Then, we have
\begin{eqnarray}
\frac{d\omega_1}{dr_1}=-\beta\gamma
\omega_1\frac{d\psi'_1}{dr_1}=\beta\gamma\omega_1
(\Omega(r_1)-\Omega_L)r_1.\label{dc15}
\end{eqnarray}
Substituting this relation in Eq. (\ref{dc11}), using the $\delta$-function
to replace $\Omega(r_1)$ by $\Omega(r)$, using ${d\psi'}/{dr}=
(-\Omega(r)+\Omega_L)r$ and comparing the resulting expression with
Eq. (\ref{d11}), we finally find that
\begin{eqnarray}
V_{r}^{pol}=-\beta\gamma D\frac{d\psi'}{dr}. \label{dc16}
\end{eqnarray}
The drift is perpendicular to the relative mean field velocity and the drift
coefficient (mobility) is given by a sort of Einstein relation
$\xi=D\beta\gamma$. We note that the drift coefficient and the
diffusion coefficient depend on the position and we recall that the
temperature is negative in cases of physical interest
\cite{onsager}. We also emphasize that the Einstein relation is valid for
the drift $V_{r}^{pol}$ due to the polarization only, not for the total
drift (\ref{dc13}). We do not have this subtlety for the usual
Brownian motion where the diffusion coefficient is constant.

If we now consider a bath with a monotonic profile of angular
velocity, using the same arguments as in Sec. \ref{sec_d}, we find that 
Eq. (\ref{dc11}) reduces to
\begin{eqnarray}
V_{r}^{pol}=2\pi^{2}\gamma\frac{\chi(r,r)}{|\Sigma(r)|}\frac{d\omega}{dr}(r).\label{dc17}
\end{eqnarray}
For the potential (\ref{pot10}), using Eq. (\ref{d13}), we find that
\begin{eqnarray}
V_{r}^{pol}=\frac{\gamma}{4}\ln\Lambda
\frac{1}{|\Sigma(r)|}\frac{d\omega}{dr}(r).\label{dc18}
\end{eqnarray}
Due to the diverging factor $\ln \Lambda\sim \frac{1}{2}\ln N$, this
expression also gives the  {\it dominant term} of the drift in the
case where the vorticity profile is non monotonic. Comparing with
Eq. (\ref{d14}), we find that the drift velocity is related to the diffusion
coefficient by the relation
\begin{eqnarray}
V_{r}^{pol}=D\frac{d\ln\omega}{dr}. \label{dc19}
\end{eqnarray}
This expression generalizes Eq. (\ref{dc16}) for a bath that is out-of-equilibrium.

\subsection{Collective effects}\label{sec_co}

As explained previously, one specificity of our approach is to develop
a formalism that allows to describe flows that are not necessarily
axisymmetric. However, its main drawback is to ignore collective
effects. In the case of axisymmetric flows, these collective effects
can be taken into account as in the study of Schecter \& Dubin
\cite{schecter}. In this section, we briefly indicate how the
preceding results can be generalized to account for these collective
effects.

For axisymmetric flows, Eqs. (\ref{dc1}) and (\ref{dc3}) can be written (we consider here the usual situation where the potential of interaction between vortices is solution of the Poisson equation):
\begin{eqnarray}
\frac{\partial\delta\omega}{\partial t}+\Omega(r,t)\frac{\partial\delta\omega}{\partial\theta}+\frac{1}{r}\frac{\partial\delta\psi}{\partial\theta}\frac{\partial\omega}{\partial r}=0,
\label{co1}
\end{eqnarray}
\begin{eqnarray}
\left\lbrack \frac{\partial^{2}}{\partial r^{2}}+\frac{1}{r}\frac{\partial}{\partial r}+\frac{1}{r^{2}}\frac{\partial}{\partial\theta^{2}}\right\rbrack\delta\psi=-\delta\omega-\gamma\frac{1}{r}\delta(r-r_{P}(t))\delta(\theta-\theta_{P}(t)),
\label{co2}
\end{eqnarray}
where $(r_{P},\theta_{P})$ are the coordinates of the test vortex.
These equations can be solved by taking the Laplace-Fourier transform of $\delta\omega$ and $\delta\psi$. Returning to physical space, the perturbed stream function can finally be written \cite{schecter}:
\begin{eqnarray}
\delta\psi(r,\theta,t)=\gamma\sum_{m}e^{im(\theta-\theta_{P})}\hat{U}_{m}(r,r_{P}),
\label{co3}
\end{eqnarray}
where
\begin{eqnarray}
\hat{U}_{m}(r,r_{P})=-\frac{1}{4\pi^{2}i}\int_{\alpha-i\infty}^{\alpha+i\infty}\frac{G(r,r_{P},m,s)}{s}e^{st}ds,
\label{co4}
\end{eqnarray}
where $G$ is the Green function solution of
\begin{eqnarray}
\left\lbrack \frac{\partial^{2}}{\partial r^{2}}+\frac{1}{r}\frac{\partial}{\partial r}-\frac{m^{2}}{r^{2}}-\frac{im}{s+im\Omega(r)}\frac{1}{r}\frac{\partial\omega}{\partial r}\right\rbrack G(r,r_{P},m,s)=\frac{\delta(r-r_{P})}{r}. 
\label{co5}
\end{eqnarray}
Therefore, when we take into account collective effects, the radial velocity created by point vortex $1$ on point vortex $0$ (say) is given by
\begin{eqnarray}
{V}_{r}(1\rightarrow 0)=\frac{\gamma}{r}\sum_{m} 
im e^{i m ({\theta}-{\theta}_{1})}\hat{U}_{m}(r,r_{1}).
\label{co6}
\end{eqnarray}
Neglecting collective effects amounts to neglecting the last term in
brackets in Eq. (\ref{co5}). It then reduces to the usual Poisson
equation where the vorticity field is due to a single point
vortex. Then, $G$ is equal to $G_{bare}=-2\pi \hat{u}_{m}(r,r_{P})$ so
that $\hat{U}_{m}(r,r_{1})$ is replaced by $\hat{u}_{m}(r,r_{1})$ in
Eq. (\ref{co6}). This returns the bare velocity (\ref{d8}) created by
point vortex $1$ on point vortex $0$.

In the computation of the diffusion coefficient, we can take into
account collective effects by replacing $\hat{u}_{m}$ by $\hat{U}_{m}$
in Eq. (\ref{d10}). This yields Eqs. (\ref{d11}) and (\ref{d12})  where
$\chi(r,r_{1})$ is replaced by
\begin{eqnarray}
\chi(r,r_1)=\sum_m {|m|}|\hat{U}_m(r,r_{1})|^2. \label{coa}
\end{eqnarray}
In fact, for $r=r_{1}$, the series diverges for large $m$ indicating
that the main contribution to the diffusion coefficient is due to
close interactions, justifying a local approximation. We thus
qualitativelty understand that collective effects will be
negligible. For large $m$, we can replace $|\hat{U}_m(r,r)|^2$ by
$|\hat{u}_m(r,r)|^2$ returning the result (\ref{d14}).

In the computation of the drift, we can take into account collective
effects as follows. The velocity created in $0$ by the
introduction of the test vortex is
\begin{eqnarray}
{V}_{r}(P\rightarrow 0)=\frac{\gamma}{r}\sum_{m}im\hat{U}_m(r,r_{P})
e^{im(\theta-\theta_{P})}. \label{co7}
\end{eqnarray}
The bare velocity due to the test vortex is
\begin{eqnarray}
{V}_{r}(P\rightarrow 0)=\frac{\gamma}{r}\sum_{m}im\hat{u}_m(r,r_{P})
e^{im(\theta-\theta_{P})}. \label{co8}
\end{eqnarray}
If we subtract Eq. (\ref{co8}) from Eq. (\ref{co7}), we get the
velocity created in $0$ by the perturbation of the distribution of the
field vortices caused by the introduction of the test
vortex. Evaluating this velocity at the position of the test vortex,
we obtain the drift experienced by the test vortex due to the
polarization process
\begin{eqnarray}
{V}_{r}^{pol}=\frac{\gamma}{r}\sum_{m}im(\hat{U}_m(r,r)-\hat{u}_m(r,r)). \label{co9}
\end{eqnarray}
This can also be written
\begin{eqnarray}
{V}_{r}^{pol}=-\frac{\gamma}{r}\sum_{m}m \ {\rm
Im}\left\lbrack \hat{U}_m(r,r)\right\rbrack. \label{co10}
\end{eqnarray}
The series diverges for $m\rightarrow +\infty$. If we replace ${\rm
Im}\lbrack\hat{U}_m(r,r)\rbrack$ by its asymptotic behaviour for large
$m$
\cite{schecter}, this returns Eq. (\ref{dc17}).

Finally, we conclude that, concerning the evaluation of the diffusion
coefficient and drift term, collective effects play a negligible role
since these quantities are dominated by close interactions. This gives
further justification to the approaches developed in Secs. \ref{sec_d}
and \ref{sec_dc}.

\subsection{The Fokker-Planck equation}\label{sec_f}

Assuming that the evolution is axisymmetric, the  probability density
$P({\bf r},t)=P(r,t)$ of finding the test vortex in ${\bf r}$ at time
$t$ is governed by a Fokker-Planck equation of the form
\begin{equation}
\label{f1} {\partial P\over\partial t}={1\over
2r}{\partial\over\partial r}\biggl\lbrack r{\partial\over\partial
r}\biggl  ({\langle (\Delta r)^{2}\rangle\over \Delta t}P\biggr
)\biggr\rbrack-{1\over r}{\partial\over\partial r}\biggl (rP{\langle
\Delta r\rangle\over \Delta t}\biggr ).
\end{equation}
This Fokker-Planck approach assumes that the stochastic process is
markovian  which is a good approximation in our case, as we have
already indicated. It also assumes that the
higher  order moments of the increment of radial position $\Delta r$
play a negligible role. This is indeed the case in the $N\rightarrow
 +\infty$ limit that we consider since they are of order $O(N^{-2})$
 or smaller. At order $O(N^{-1})$, we have found that the second (diffusion) and first (drift) moments
 of the radial increments of position of the
test vortex are given  by
\begin{equation}
\label{f2}{\langle (\Delta r)^{2}\rangle\over 2 \Delta t}=D, \qquad
{\langle \Delta r\rangle\over \Delta t}={\partial D\over\partial
r}+\eta,
\end{equation}
with
\begin{eqnarray}
D=\frac{2\pi^{2}\gamma}{r^{2}}\int_{0}^{+\infty}r_{1}dr_{1}\chi(r,r_{1})\delta(\Delta\Omega)\omega({r}_{1}),\label{f3}
\end{eqnarray}
\begin{eqnarray}
\eta\equiv
V_{r}^{pol}=\frac{2\pi^{2}\gamma}{r}\int_{0}^{+\infty}dr_{1}\chi(r,r_{1})\delta(\Delta\Omega)\frac{d\omega}{d r} ({r}_{1}).\label{f4}
\end{eqnarray}
The Fokker-Planck equation (\ref{f1}) can be written in the alternative
form
\begin{equation}
\label{f5}{\partial P\over\partial t}=\frac{1}{r}{\partial\over\partial r}\biggl\lbrack r\biggl (D{\partial P\over\partial r}-P\eta\biggr )\biggr\rbrack.
\end{equation}
The two expressions (\ref{f1}) and (\ref{f5}) have their own
interest. The expression (\ref{f1}) where the diffusion coefficient is
placed after the second derivative $\partial^{2}(DP)$ involves the
total drift $V_r^{drift}=\langle \Delta r\rangle/\Delta t$ and the
expression (\ref{f5}) where the diffusion coefficient is placed
between the derivatives $\partial D\partial P$ isolates the part of
the drift $\eta=V_{r}^{polar}$ due to the polarization. This
alternative form (\ref{f5}) has therefore a clear physical
interpretation. Inserting the expressions (\ref{f3}) and (\ref{f4}) of
the diffusion coefficient and drift term in Eq. (\ref{f5}), we obtain
\begin{eqnarray}
\frac{\partial P}{\partial t}=2\pi^2 \gamma \frac{1}{r}
\frac{\partial}{\partial r} \int_0^{+\infty} r_1 dr_1 \chi(r,r_1)
\delta(\Omega-\Omega_1) \left (\frac{1}{r}\frac{\partial}{\partial
r}-\frac{1}{r_1}\frac{\partial}{\partial
 r_1}\right )P(r,t)\omega(r_{1}). \label{f6}
\end{eqnarray}
For a thermal bath, using Eq. (\ref{dc16}), the Fokker-Planck equation
(\ref{f5}) can be written
\begin{equation}
\label{f7}{\partial P\over\partial
t}=\frac{1}{r}{\partial\over\partial r}\biggl\lbrack r D(r)\biggl
({\partial P\over\partial r}+\beta\gamma P\frac{d\psi'}{dr}\biggr
)\biggr\rbrack,
\end{equation}
where $D(r)$ is given by Eq. (\ref{f3}). For a steady bath with a
monotonic vorticity profile, using Eq. (\ref{dc19}), the Fokker-Planck
equation (\ref{f5}) can be written
\begin{equation}
\label{f8}{\partial P\over\partial
t}=\frac{1}{r}{\partial\over\partial r}\biggl\lbrack r D(r)\biggl
({\partial P\over\partial r}-P\frac{d\ln\omega}{dr}\biggr
)\biggr\rbrack,
\end{equation}
where $D(r)$ is given by Eq. (\ref{d14}). These Fokker-Planck
equations have been studied in detail by Chavanis \& Lemou
\cite{chavlemou} for different types of bath distribution.  The
distribution of the test vortex relaxes to the distribution of the
bath on a typical timescale $(N/\ln N)t_D$ but the relaxation process
is very peculiar and differs from the usual exponential relaxation. In
particular, the evolution of the front profile in the tail of the
distribution is very slow (logarithmic) and the temporal correlation
function $\langle r(0)r(t)\rangle$ decreases like $\ln t/t$ (for a
thermal bath). This is due to the rapid decay of the diffusion
coefficient $D(r)$, like in the HMF model \cite{bd}.

In our previous papers, we have obtained Eq. (\ref{f6}) directly from
the projection operator formalism (see Sec. 4.1 of
\cite{chavlemou}). This amounts, in the kinetic equation (\ref{ma8}), to replacing $\omega(r,t)$ by the distribution  $P(r,t)$ of the test vortex
and $\omega(r_1,t)$ by the {\it static} distribution $\omega(r_1)$ of
the field vortices. This procedure transforms an integrodifferential
equation (\ref{ma8}) into a differential equation (\ref{f6}). Then,
the expressions (\ref{f2})-(\ref{f4}) of the diffusion and drift terms
were obtained by identifying Eq. (\ref{f6}) with the Fokker-Planck
equation (\ref{f1}). In the present paper, we have proceeded the other
way round by first determining the moments (\ref{f2})-(\ref{f4}) in
Secs. \ref{sec_d} and \ref{sec_dc}, then inserting them in the
Fokker-Planck equation (\ref{f1}). Although this procedure may appear
more logical in some sense, the other approach based on the projection
operator formalism is more powerful because it allows one to obtain
more general equations that are non markovian and that relax the
hypothesis of axisymmetry as discussed in the next section.

\subsection{More general kinetic equations}\label{sec_mg}

It is instructive to compare the Fokker-Planck equation (\ref{f6})
with the more general equation obtained from the projection operator
formalism
\begin{eqnarray}
\frac{\partial P}{\partial t}+\langle {\bf V}\rangle \nabla P=\frac{\partial}{\partial r^{\mu}}\int_{0}^{t}d\tau\int d{\bf r}_{1}{V}^{\mu}(1\rightarrow 0)
G(t,t-\tau)\nonumber\\
\times \biggl\lbrace {\cal V}^{\nu}(1\rightarrow 0)\frac{\partial}{\partial r^{\nu}}
+ {\cal V}^{\nu}(0\rightarrow 1)\frac{\partial}{\partial r_{1}^{\nu}}\biggr\rbrace P({\bf r},t-\tau)\frac{\omega}{\gamma}({\bf r}_{1}).
\label{mg1}
\end{eqnarray}
This equation can be obtained from Eq. (\ref{s1}), by replacing
$\omega(r,t)$ by $P(r,t)$ and $\omega(r_1,t)$ by $\omega(r_1)$. This
is a sort of generalized ``Fokker-Planck'' equation involving a term
of ``diffusion'' and a term of ``friction''. However, strictly
speaking, Eq. (\ref{mg1}) is not a Fokker-Planck equation because it
is non-Markovian. We also note that the ``diffusion'' term appears as
a complicated time integral of the velocity correlation function
involving $P({\bf r},t-\tau)$. This can be seen as a generalization of
the Kubo formula (\ref{d5}). Similarly the ``drift'' is a
generalization of the expression obtained in (\ref{dc6}) with a more
complicated time integral. If we consider a thermal bath where the distribution of the field vortices is the Boltzmann distribution, we get
\begin{eqnarray}
\frac{\partial P}{\partial t}+\langle {\bf V}\rangle \nabla P=\frac{\partial}{\partial r^{\mu}}\int_{0}^{t}d\tau\int d{\bf r}_{1}{V}^{\mu}(1\rightarrow 0)
G(t,t-\tau)\nonumber\\
\times \biggl\lbrace {\cal V}(1\rightarrow 0)\cdot \nabla
-\beta\gamma {\cal V}(0\rightarrow 1)\cdot \nabla\psi({\bf r}_{1})\biggr\rbrace P({\bf r},t-\tau)\frac{\omega}{\gamma}({\bf r}_{1}).
\label{mg1bis}
\end{eqnarray} 
If we come back to Eq. (\ref{mg1}), make a Markovian approximation and
extend the time integral to infinity, we obtain
\begin{eqnarray}
\frac{\partial P}{\partial t}+\langle {\bf V}\rangle \nabla P=\frac{\partial}{\partial r^{\mu}}\int_{0}^{+\infty}d\tau\int d{\bf r}_{1}{V}^{\mu}(1\rightarrow 0)
G(t,t-\tau)\nonumber\\
\times \biggl\lbrace {\cal V}^{\nu}(1\rightarrow 0)\frac{\partial}{\partial r^{\nu}}
+ {\cal V}^{\nu}(0\rightarrow 1)\frac{\partial}{\partial r_{1}^{\nu}}\biggr\rbrace P({\bf r},t)\frac{\omega}{\gamma}({\bf r}_{1}),
\label{mg1b}
\end{eqnarray} 
where we recall that the coordinates appearing after the Greenian must
be viewed as explicit functions of time, i.e.  ${\bf r}_{i}={\bf r}_{i}(t-\tau)$. On the
other hand, for an axisymmetric system, using the relation
(\ref{pot5}) and $r_{i}(t-\tau)=r_{i}(t)$ and $\theta_{i}(t-\tau)=\theta_{i}(t)-\Omega(r_{i}(t))\tau$, Eq. (\ref{mg1}) takes the
simplest form
\begin{eqnarray}
\frac{\partial P}{\partial t}=\frac{1}{r}\frac{\partial}{\partial
r}r\int_{0}^{t}d\tau \int_{0}^{2\pi}d\theta_{1}\int_{0}^{+\infty} r
r_{1} d{r}_{1}{V}_{r}(1\rightarrow 0,t)\nonumber\\
\times {V}_{r}(1\rightarrow 0,t-\tau)\biggl
(\frac{1}{r}\frac{\partial}{\partial r}
-\frac{1}{r_{1}}\frac{\partial}{\partial r_{1}}\biggr )
{P}(r,t-\tau)\frac{\omega}{\gamma}({r}_{1}). \label{mg2}
\end{eqnarray}
If we make a Markovian approximation ${P}(r,t-\tau)\simeq P(r,t)$
and extend the time integral to infinity, we get
\begin{eqnarray}
\frac{\partial P}{\partial t}=\frac{1}{r}\frac{\partial}{\partial r}r\int_{0}^{+\infty}d\tau
\int_{0}^{2\pi}d\theta_{1}\int_{0}^{+\infty} r r_{1} d{r}_{1}{V}_{r}(1\rightarrow 0,t)\nonumber\\
\times {V}_{r}(1\rightarrow 0,t-\tau)\biggl
(\frac{1}{r}\frac{\partial}{\partial r}
-\frac{1}{r_{1}}\frac{\partial}{\partial r_{1}}\biggr )
{P}(r,t)\frac{\omega}{\gamma}({r}_{1}). \label{mg3}
\end{eqnarray}
This is a Fokker-Planck equation which can be put in the form (\ref{f5})
with a diffusion coefficient
\begin{eqnarray}
D=\int_{0}^{+\infty} d\tau
\int_{0}^{2\pi}d\theta_{1}\int_{0}^{+\infty} r_{1}d{r}_{1}
{V}_{r}(1\rightarrow 0,t){V}_{r}(1\rightarrow
0,t-\tau)\frac{\omega}{\gamma}({r}_{1}), \label{mg4}
\end{eqnarray}
and a drift term due to the polarization
\begin{eqnarray}
\eta=\frac{1}{\gamma}\int_{0}^{+\infty}d\tau\int_{0}^{2\pi}d\theta_{1}\int_{0}^{+\infty}rdr_{1}{V}_{r}(1\rightarrow
0){V}_{r}(1\rightarrow 0,t-\tau)\frac{d \omega}{d r}
({r}_{1}). \label{mg5}
\end{eqnarray}
These expressions agree with Eqs. (\ref{d7}) and (\ref{dc10}) obtained
directly from the equations of motion. After integration on $\tau$ and
$\theta_1$, we recover the Fokker-Planck equation (\ref{f6}) with the
expressions (\ref{f3}) and (\ref{f4}) of the diffusion coefficient and
drift term.

\section{Conclusion}\label{sec_conc}

In this paper, we have developed the kinetic theory of point vortices
in two-dimensional hydrodynamics initiated in \cite{kin}. Point
vortices provide a fundamental example of systems with long-range
interactions
\cite{houches} which deserves a particular attention. We have shown
that the main features of the kinetic theory: kinetic equation
describing the evolution of the system as a whole, diffusion
coefficient, drift term, Fokker-Planck equation describing the
evolution of a test particle in a bath... could be obtained from a
simpler formalism than the one developed in our previous papers
\cite{preR,kin}. This clarifies the argumentation and
delineates the domain of validity of the theory. We have given general
equations that are valid for flows that are not necessarily
axisymmetric nor markovian. A limitation of our approach is to neglect
collective effects. These effects  have been taken into account in
\cite{dn,schecter} for axisymmetric flows. In plasma physics, collective effects are important because they lead to Debye shielding and 
regularize the logarithmic divergence at large scales that appears in
the Landau equation (as shown by Lenard \cite{lenard} and Balescu
\cite{balescu}). For point vortices, their influence seems less crucial since
the kinetic equation (\ref{ma10}) derived by neglecting collective
terms does not present any divergence. In addition, concerning the
expressions of the diffusion coefficient and drift term, we have
indicated that collective effects have a negligible contribution
because the diffusion coefficient and the drift velocity are dominated
by local interactions. In future works, we plan to study in more
detail the kinetic equations given in this paper. This project has
been initiated in \cite{chavlemou}. The kinetic
theory could be used to interprete the numerical simulations of point
vortices in 2D hydrodynamics \cite{kw1,kw2} or the experiments of
non-neutral plasmas under a strong magnetic field (leading to quasi
stationary states, vortex crystals,...) \cite{fine}. In agreement with
the kinetic theory, these systems exhibit a violent collisionless
relaxation and a slow collisional evolution. The collisionless
relaxation is based on the 2D Euler equation and the evolution of the
coarse-grained vorticity is described by Eq. (\ref{v3}). The
collisional evolution was the main object of interest of the present
paper. It is described by a general kinetic equation of the form
(\ref{gen6}) that simplifies in Eq. (\ref{ma10}) for axisymmetric
flows.

\appendix

\section{The potential of interaction}\label{sec_pot}

The velocity of the $i$-th vortex is produced by the other vortices
according to the relation
\begin{eqnarray}
{\bf V}_i=-\frac{1}{\gamma}{\bf z}\times\nabla_{i} H=\sum_{j\neq
i}{\bf V}(j\rightarrow i),\label{pot1}
\end{eqnarray}
where $H$ is the Hamiltonian (\ref{b2}).  The velocity created by point
vortex $j$ on point vortex $i$ is
\begin{eqnarray}
{\bf V}(j\rightarrow i)=-\gamma{\bf z}\times \frac{\partial
u_{ij}}{\partial {\bf r}_i}.\label{pot2}
\end{eqnarray}
Introducing a system of polar coordinates to localize the point
vortices, the radial component in the direction of ${\bf r}_1$ of
the velocity created by point vortex $2$ on point vortex $1$ is
\begin{eqnarray}
{V}_{r_1}(2\rightarrow 1)=\frac{\gamma}{r_1}\frac{\partial
u_{12}}{\partial \theta_1}.\label{pot3}
\end{eqnarray}
In an infinite domain, the potential of interaction can be written
\begin{eqnarray}
u_{12}=u(\sqrt{r_1^2+r_2^2-2r_1 r_2 \cos\phi})\equiv
u(r_1,r_2,\phi),\label{pot4}
\end{eqnarray}
where $\phi=\theta_1-\theta_2$. This implies that
\begin{eqnarray}
V_{r_2}(1\rightarrow 2)=-\frac{r_1}{r_2}V_{r_1}(2\rightarrow
1).\label{pot5}
\end{eqnarray}
This relation results from the conservation of the angular momentum
(see Appendix D of \cite{kin}) and remains valid in a bounded circular
domain. Since the function $u(r_1,r_2,\phi)$ is periodic with period
$2\pi$, it can be decomposed in Fourier series. Thus,
\begin{eqnarray}
u(r_1,r_2,\phi)=\sum_m e^{im\phi}\hat{u}_m(r_1,r_2),\label{pot6}
\end{eqnarray}
with
\begin{eqnarray}
\hat{u}_m(r_1,r_2)=\frac{1}{2\pi}\int_0^{2\pi}
\cos(n\phi)u(r_1,r_2,\phi)d\phi. \label{pot7}
\end{eqnarray}
Using the decomposition (\ref{pot6}), we find that Eq. (\ref{pot3})
can be rewritten in the form
\begin{eqnarray}
{V}_{r_1}(2\rightarrow 1)=i\gamma \frac{1}{r_1}\sum_{m} m
\hat{u}_{m}(r_1,r_2)e^{i m (\theta_1-\theta_{2})}. \label{pot8}
\end{eqnarray}
The usual potential of interaction between point vortices is
solution of the Poisson equation
\begin{eqnarray}
\Delta u=-\delta({\bf r}). \label{pot9}
\end{eqnarray}
In an infinite domain, we have
\begin{eqnarray}
u_{12}=-\frac{1}{2\pi}\ln |{\bf r}_1-{\bf r}_2|.\label{pot10}
\end{eqnarray}
The Fourier transform of $u(\phi)$ can be easily obtained by using, e.g., 
the identities given in Appendix E1 of \cite{kin}. We find
\begin{eqnarray}
\hat{u}_m(r_1,r_2)=\frac{1}{4\pi |m|}\left (\frac{r_<}{r_>}\right
)^{|m|}, \qquad (m\neq 0)\label{pot11}
\end{eqnarray}
\begin{eqnarray}
\hat{u}_0(r_1,r_2)=-\frac{1}{2\pi}\ln r_>,\label{pot12}
\end{eqnarray}
where $r_{>}$ (resp. $r_{<}$) is the largest (resp. smallest) of
$r_{1}$ and $r_{2}$.  In that case, the function defined in Eq. (\ref{ma9})
takes the explicit form
\begin{eqnarray}
\chi(r_1,r_2)=\frac{1}{8\pi^2}\sum_{m=1}^{+\infty} \frac{1}{m} \left
(\frac{r_<}{r_>}\right )^{2m}=-\frac{1}{8\pi^2}\ln\left\lbrack
1-\left (\frac{r_<}{r_>}\right )^{2}\right\rbrack. \label{pot13}
\end{eqnarray}
When the point vortices are confined within a circular box of radius $R$, the
potential of interaction is
\begin{eqnarray}
u_{12}=-\frac{1}{2\pi}\ln |{\bf r}_1-{\bf r}_2|+\frac{1}{2\pi}\ln
\left|{\bf r}_1-\frac{R^2}{r_2^2}{\bf r}_2\right|,\label{pot14}
\end{eqnarray}
and its Fourier transform is given by
\begin{eqnarray}
\hat{u}_m=\frac{1}{4\pi |m|}\left (\frac{r_<}{r_>}\right
)^{|m|}\left \lbrack 1-\left (\frac{r_{>}}{R}\right
)^{2|m|}\right\rbrack, \qquad (m\neq 0)\label{pot15}
\end{eqnarray}
\begin{eqnarray}
\hat{u}_0(r_1,r_2)=\frac{1}{2\pi}\ln \left (\frac{R^2}{r_2
r_>}\right ).\label{pot16}
\end{eqnarray}
In that case, the function defined in Eq. (\ref{ma9}) takes the explicit
form
\begin{eqnarray}
\chi(r_1,r_2)=-\frac{1}{8\pi^2}\ln\left\lbrack 1-\left
(\frac{r_<}{r_>}\right
)^{2}\right\rbrack+\frac{1}{4\pi^2}\ln\left\lbrack 1-\left
(\frac{r_<}{R}\right
)^{2}\right\rbrack-\frac{1}{8\pi^2}\ln\left\lbrack 1-\left
(\frac{r_<r_>}{R^2}\right )^{2}\right\rbrack.\label{pot17}
\end{eqnarray}
Finally, in  the QG model, the potential of interaction is solution
of an equation of the form
\begin{eqnarray} \Delta u-\frac{1}{L^2}u=-\delta({\bf
r}),\label{pot18}
\end{eqnarray}
where $L$ is the Rossby radius. In an infinite domain, we obtain
\begin{eqnarray}
u_{12}=\frac{1}{2\pi}K_0\left (\frac{|{\bf r}_1-{\bf r}_2|}{L}\right
).\label{pot19}
\end{eqnarray}

\section{A heuristic kinetic equation}\label{sec_heur}

In \cite{kin} and in Sec. \ref{sec_gen}, we have derived a general
kinetic equation (\ref{gen6}) or (\ref{s1b}) that is valid at order
$O(1/N)$. This equation can be simplified for axisymmetric flows
(leading to Eq. (\ref{ma10})) and uni-directional flows (leading to
Eq. (135) of
\cite{kin}). In \cite{kin}, we have heuristically proposed a
generalized kinetic equation (137) that encompasses both the
axisymmetric and unidirectional forms. This equation is not exact, so
it cannot be obtained rigorously from Eq. (\ref{gen6}). Yet, it
possesses interesting properties (conservation of $E$, $\Gamma$, $L$,
$P$ and $H$-theorem $\dot S\ge 0$) so it can be useful. For
axisymmetric and unidirectional flows, it does not exactly reduce to
Eq.  (\ref{ma10}) and Eq.  (135) of \cite{kin}, but it has a similar
structure so that the disagreement is not too severe. In this
Appendix, we try to justify this equation but we stress that, since
this equation is not exact, some approximations are necessarily
un-controlled.

First, assuming that the decorrelation time is {\it extremely} short
(which does not need to be the case) and that ${\cal V}\simeq {\bf
V}$, we replace Eq. (\ref{s1b}) by
\begin{eqnarray}
\frac{\partial \omega_1}{\partial t}+\frac{N-1}{N}\langle {\bf
V}\rangle_{1} {\partial \omega\over \partial {\bf r}_{1}}
=\frac{\partial}{\partial {r}_1^{\mu}}\int_0^{+\infty} d\tau \int d{\bf
r}_{2} {V}^{\mu}(2\rightarrow1,t){V}^{\nu}(2\rightarrow 1,t-\tau)\nonumber\\
\times\left (
{\partial\over\partial { r}_{1}^{\nu}}-{\partial\over\partial {r}_{2}^{\nu}}\right )\omega({\bf
r}_1,t)\frac{\omega}{\gamma}({\bf r}_2,t), \label{heur1}
\end{eqnarray}
where, now, the vorticity and vorticity gradient are evaluated at
${\bf r}_{1}={\bf r}_{1}(t)$ and ${\bf r}_{2}={\bf
r}_{2}(t)$. Introducing the Fourier transform of the velocity created
by point vortex $2$ on point vortex $1$:
\begin{eqnarray}
{\bf V}(2\rightarrow 1,t)=-i\gamma\int{\bf k}_{\perp}\hat{u}(k)e^{i{\bf k}({\bf r}_{1}(t)-{\bf r}_{2}(t))}d{\bf k},
 \label{heur2}
\end{eqnarray}
and making a linear trajectory approximation ${\bf r}_{i}(t-\tau)\simeq {\bf r}_{i}(t)-\langle {\bf V}\rangle({\bf r}_{i},t)\tau$, we get
\begin{eqnarray}
\frac{\partial \omega_1}{\partial t}+\frac{N-1}{N}\langle {\bf
V}\rangle_{1} {\partial \omega\over \partial {\bf r}_{1}}
=-\gamma^{2}\frac{\partial}{\partial {r}_1^{\mu}}\int_0^{+\infty} d\tau \int d{\bf
r}_{2}\int d{\bf k}d{\bf k}' k_{\perp}^{\mu}k_{\perp}^{'\nu}\hat{u}(k)\hat{u}(k')e^{-i({\bf k}+{\bf k}')\cdot {\mb\xi}}e^{i{\bf k}'\cdot {\bf v}\tau} \nonumber\\
\times\left (
{\partial\over\partial { r}_{1}^{\nu}}-{\partial\over\partial {r}_{2}^{\nu}}\right )\omega({\bf
r}_1,t)\frac{\omega}{\gamma}({\bf r}_2,t),\label{heur3}
\end{eqnarray}
where ${\mb\xi}={\bf r}_{2}-{\bf r}_{1}$ and ${\bf v}=\langle {\bf
V}\rangle({\bf r}_{2},t)-\langle {\bf V}\rangle({\bf r}_{1},t)$. The
linear trajectory approximation is clearly not justified for an
axisymmetric flow (since the point vortices follow circular trajectories as considered in Sec. \ref{sec_ma}) so
it again relies on the (un-controlled) hypothesis that the
decorrelation time is extremely short. Integrating on time, we obtain
\begin{eqnarray}
\frac{\partial \omega_1}{\partial t}+\frac{N-1}{N}\langle {\bf
V}\rangle_{1} {\partial \omega\over \partial {\bf r}_{1}}
=-\pi\gamma^{2}\frac{\partial}{\partial {r}_1^{\mu}} \int d{\bf
r}_{2}\int d{\bf k}d{\bf k}' k_{\perp}^{\mu}k_{\perp}^{'\nu}\hat{u}(k)\hat{u}(k')e^{-i({\bf k}+{\bf k}')\cdot {\mb\xi}}\delta({\bf k}'\cdot {\bf v}) \nonumber\\
\times\left (
{\partial\over\partial { r}_{1}^{\nu}}-{\partial\over\partial {r}_{2}^{\nu}}\right )\omega({\bf
r}_1,t)\frac{\omega}{\gamma}({\bf r}_2,t).\label{heur4}
\end{eqnarray}
Now, using the heuristic argument that ${\bf k}'\cdot
{\mb\xi}\simeq 1$, i.e. ${\bf k}'\simeq {\mb\xi}/\xi^{2}$ (due to the
exponential $e^{-i{\bf k}'\cdot {\mb\xi}}$), we make the rough
substitution $\delta ({\bf k}'\cdot {\bf v})\rightarrow \lambda
\delta(({\mb \xi}/\xi^{2})\cdot {\bf v})=\lambda
\xi^{2}\delta({\mb \xi}\cdot {\bf v})$, where $\lambda$ is a constant of order unity, in Eq. (\ref{heur4}). Then,
using Eq. (\ref{heur2}), we can rewrite Eq. (\ref{heur4}) in the form
\begin{eqnarray}
\frac{\partial \omega_1}{\partial t}+\frac{N-1}{N}\langle {\bf
V}\rangle_{1} {\partial \omega\over \partial {\bf r}_{1}}
=\lambda\pi\frac{\partial}{\partial {r}_1^{\mu}} \int d{\bf
r}_{2}V^{\mu}(2\rightarrow 1,t)V^{\nu}(2\rightarrow 1,t)\xi^{2}\delta({\mb \xi}\cdot {\bf v})\nonumber\\
\times\left (
{\partial\over\partial { r}_{1}^{\nu}}-{\partial\over\partial {r}_{2}^{\nu}}\right )\omega({\bf
r}_1,t)\frac{\omega}{\gamma}({\bf r}_2,t).\label{heur5}
\end{eqnarray}
Finally, using ${\bf V}(2\rightarrow 1)=-(\gamma/2\pi){\mb\xi}_{\perp}/\xi^{2}$, we obtain
\begin{eqnarray}
\frac{\partial \omega_1}{\partial t}+\frac{N-1}{N}\langle {\bf
V}\rangle_{1} {\partial \omega\over \partial {\bf r}_{1}}
=\frac{\gamma}{8}\frac{\partial}{\partial {r}_1^{\mu}} \int d{\bf
r}_{2}\frac{\xi^{2}\delta^{\mu\nu}-\xi^{\mu}\xi^{\nu}}{\xi^{2}}\delta({\mb \xi}\cdot {\bf v})\left (
{\partial\over\partial { r}_{1}^{\nu}}-{\partial\over\partial {r}_{2}^{\nu}}\right )\omega({\bf
r}_1,t){\omega}({\bf r}_2,t).\label{heur6}
\end{eqnarray}
This is Eq. (137) of \cite{kin}. The constant $\lambda$ has been
determined so that Eq. (\ref{heur6}) reproduces at best the exact
equations (Eq. (\ref{ma10}) and Eq. (135) of
\cite{kin}) obtained for axisymmetric and unidirectional flows (in particular the expressions of the drift and
diffusion in the corresponding Fokker-Planck equation, see
\cite{kin}). This yields $\lambda=\pi/2$ which is of order unity 
as expected. The previous arguments give some ``justification'' to
Eq. (137) of
\cite{kin} although we again stress that this equation is not exact so
it is obtained from un-controlled approximations that are not really
justified.

\section{Drift term for axisymmetric systems}\label{sec_da}

For an axisymmetric flow, the linearized equation (\ref{dc7}) for the
perturbation becomes
\begin{eqnarray}
\frac{\partial\delta\omega}{\partial t}+u_{\theta}\frac{1}{r}\frac{\partial \delta\omega}{\partial\theta}=-{V}_{r}(P\rightarrow 0)\frac{d \omega}{d r}.
\label{da1}
\end{eqnarray}
Introducing the angular velocity and the potential of interaction, it
can be rewritten
\begin{eqnarray}
\frac{\partial\delta\omega}{\partial t}+\Omega(r)\frac{\partial \delta\omega}{\partial\theta}=-\frac{\gamma}{r}\frac{\partial u_{OP}}{\partial \theta}\frac{d \omega}{d r}.
\label{da2}
\end{eqnarray}
Taking the Fourier transform of this expression with respect to the
angular variable, we obtain
\begin{eqnarray}
\frac{d\delta\hat{\omega}_{m}}{\partial t}+im\Omega\delta\hat{\omega}_{m}=-\frac{\gamma}{r}im \hat{u}_{m}(r,r_{P})e^{-im\theta_{P}}\frac{d\omega}{dr}.
\label{da3}
\end{eqnarray}
Integrating this first order differential equation with respect to
time, we find that
\begin{eqnarray}
\delta\hat{\omega}_{m}(t)=-\int_{0}^{t}d\tau e^{-im\Omega\tau} \frac{\gamma}{r}im \hat{u}_{m}(r,r_{P})e^{-im\theta_{P}(t-\tau)}\frac{d\omega}{dr}.
\label{da4}
\end{eqnarray}
To leading order in $N\rightarrow +\infty$, the test particle is advected by the mean flow so that $\theta_{P}(t-\tau)=\theta_{P}-\Omega_{P}\tau$. Extending the time integral to $+\infty$, we obtain
\begin{eqnarray}
\delta\hat{\omega}_{m}(t)=-\int_{0}^{+\infty}d\tau e^{-im(\Omega-\Omega_{P})\tau} \frac{\gamma}{r}im \hat{u}_{m}(r,r_{P})e^{-im\theta_{P}}\frac{d\omega}{dr}.
\label{da5}
\end{eqnarray}
The integral on time $\tau$ can be easily calculated yielding
\begin{eqnarray}
\delta\hat{\omega}_{m}(t)=-\pi \delta\lbrack m(\Omega-\Omega_{P})\rbrack \frac{\gamma}{r}im \hat{u}_{m}(r,r_{P})e^{-im\theta_{P}}\frac{d\omega}{d r}.
\label{da6}
\end{eqnarray}
The drift velocity experienced by the test particle is
\begin{eqnarray}
V_{r}^{pol}=\frac{1}{\gamma}\int V_{r}(1\rightarrow P)\delta\omega_{1}(t)d{\bf r}_{1},
\label{da7}
\end{eqnarray}
or, more explicitly,
\begin{eqnarray}
V_{r}^{pol}=\int_{0}^{+\infty}r_{1}dr_{1}\int_{0}^{2\pi}d\theta_{1} \frac{1}{r_{P}}\frac{\partial u_{P1}}{\partial \theta_{P}}\delta\omega_{1}.
\label{da8}
\end{eqnarray}
Introducing Fourier transforms and substituting Eq. (\ref{da6}) in
Eq. (\ref{da8}), this can be rewritten
\begin{eqnarray}
V_{r}^{pol}=\frac{2\pi^{2}\gamma}{r_{P}}\int_{0}^{+\infty}dr_{1}\sum_{m}m^{2}\hat{u}_{m}(r_{1},r_{P})^{2}\delta\lbrack m(\Omega_{1}-\Omega_{P})\rbrack \frac{d\omega_{1}}{d r_{1}}.
\label{da9}
\end{eqnarray}
Finally, introducing the notation (\ref{ma9}), we obtain
\begin{eqnarray}
V_{r}^{pol}=\frac{2\pi^{2}\gamma}{r_{P}}\int_{0}^{+\infty}dr_{1}\chi(r_{1},r_{P}) \delta(\Omega_{1}-\Omega_{P})\frac{d\omega_{1}}{dr_{1}}. 
\label{da10}
\end{eqnarray}


\end{document}